\documentclass[prb,aps,superscriptaddress,twocolumn]{revtex4-1}
\usepackage{graphicx}
\usepackage{bm}
\usepackage{amsmath}
\usepackage{amsfonts}
\usepackage{amssymb}
\usepackage{epsfig}
\usepackage{color}
\usepackage{ulem}

\begin{document}


\title{Spin-Orbital Hallmarks of Unconventional Superconductors \\ Without Inversion Symmetry}



\author{Yuri Fukaya}
\affiliation{Department of Applied Physics, Nagoya University, Nagoya 464-8603, Japan}

\author{Shun Tamura}
\affiliation{Department of Applied Physics, Nagoya University, Nagoya 464-8603, Japan}

\author{Keiji Yada}
\affiliation{Department of Applied Physics, Nagoya University, Nagoya 464-8603, Japan}

\author{Yukio Tanaka}
\affiliation{Department of Applied Physics, Nagoya University, Nagoya 464-8603, Japan}

\author{Paola Gentile}
\affiliation{CNR-SPIN, I-84084 Fisciano (Salerno), Italy, c/o Universit\'a di Salerno, I-84084 Fisciano (Salerno), Italy}

\author{Mario Cuoco}
\affiliation{CNR-SPIN, I-84084 Fisciano (Salerno), Italy, c/o Universit\'a di Salerno, I-84084 Fisciano (Salerno), Italy}



\begin{abstract}
The spin-orbital polarization of superconducting excitations in momentum space is shown to provide distinctive marks of unconventional pairing in the presence of inversion symmetry breaking. 
Taking the prototypical example of an electronic system with atomic spin-orbit and orbital-Rashba couplings, we provide a general description of the spin-orbital textures and their most striking changeover moving from the normal to the superconducting state. 
We find that the variation of the spin-texture is strongly imprinted by the combination of the misalignment of spin-triplet ${\bf d}$-vector with the inversion asymmetry ${\bf g}$-vector coupling and the occurrence of superconducting nodal excitations. Remarkably, the multi-orbital character of the superconducting state allows to unveil a unique type of topological transition for the spin-winding around the nodal points. 
This finding indicates the fundamental topological relation between chiral and spin-winding in nodal superconductors.
By analogy between spin- and orbital-triplet pairing we point out how orbital polarization patterns can also be employed to assess the character of the superconducting state.
\end{abstract}

\maketitle

\section{Introduction}

The Rashba spin-orbit (SO) coupling~\cite{rashba60,bycchkov84} is the manifestation of a fundamental relativistic effect due to structural inversion symmetry breaking (ISB) that leads to spin-momentum locking with lifting of spin degeneracy and remarkable phenomena such as non-standard magnetic textures~\cite{dzyalo57,moriya60}, spin Hall~\cite{sinova04} and topological spin Hall~\cite{chang13}, Edelstein effects~\cite{edelstein90}, etc.\ \cite{manchon15}. 

Recently, it has been realized that spin-momentum locking can also occur from the ISB driven orbital polarization of electrons in solids which is, then, linked with the spin-sector by the atomic SO coupling. 
The role of spin and orbital polarization in materials has built a different view of the manifestation of ISB with respect to the conventional spin-Rashba effect, leading to the so-called orbital-driven Rashba coupling~\cite{park11}. The orbital Rashba (OR) effect can yield chiral orbital textures and orbital dependent spin-vector via the SO coupling~\cite{park11,park13,kim14,hong15,kim12,park12,park12a}.
Evidences of anomalous energy splitting and of a key role played by the orbital degree of freedom have been demonstrated on a large variety of surfaces, i.e.\ Au(111), Pb/Ag(111)~\cite{el-kareh14}, Bi/Ag(111)~\cite{schirone15}, etc.\ as well as in transition metal oxides based interfaces, i.e., LaAlO$_3$-SrTiO$_3$~\cite{King2014,Nakamura2012}.

In superconductors without inversion symmetry~\cite{Bauer12,Smidman17} the presence of non-degenerate spin- and orbital polarized electronic states is generally expected to lead to unconventional pairing, with the occurrence of spin-triplet order parameters and singlet-triplet spin mixing~\cite{Gorkov2001,Frigeri2004,Yada2009}, non-standard surface states~\cite{Lu2008,Vorontsov2008}, as well as topological phases~\cite{Tanaka2009,Sato2009, Schnyder2010,Mizuno2010,Yada2011,Sato2011,Brydon2011,Scheurer2015,Ying2017}. 

Experimental direct probes by using angle- and spin-orbital resolved photoemission spectroscopy in the normal~ \cite{Borisenko16,Johnson16,Wu17,Damascelli18}
and superconducting (SC) phase~\cite{neupane15} can be extremely useful for establishing the nature of the SC state and the underlying degree of spin-orbital entanglement or the occurrence of competing orders. 
A successful photoemission observation of Dirac-cone with spin-helical surface states at Fermi level and their modification below the superconducting critical temperature due to the gap opening has been recently demonstrated in the iron-based superconductor FeTe$_{1-x}$Se$_x$\cite{Zhang2018}.
Along this line it would be highly desirable to have distinctive detectable signatures associated with the spin-orbital polarizations to single out the nature of the SC phase. Symmetry plays a relevant role in such identification.  
For instance, skyrmionic patterns in the Brillouin zone (BZ) have been suggested as marks to make the topological order more accessible in ferromagnetic semiconductor/$s$-wave superconductor heterostructure assuming that both time-reversal (TR) and inversion symmetry is broken~\cite{schaffer}. On the other hand, the fundamental interrelation between chiral spin-orbital textures in reciprocal space and unconventional pairing solely due to ISB has not been yet fully established. 

In this paper we focus on the class of low-dimensional superconductors with TR and broken inversion symmetry. The aim is to assess how the spin-orbital texture of the SC excitations can unveil the nature of the SC state and, eventually, its topological character. 

We show that the spin-polarization pattern is generally imprinted by the relative alignment of spin-triplet ${\bf d}$-vector with the inversion asymmetry ${\bf g}$-vector coupling (Sect.\ II).
A fundamental issue emerges in nodal topological superconductors when considering the occurrence of spin-winding around the nodal points. 
To face this problem on a general ground we employ a prototypical electronic system with atomic SO and orbital-Rashba coupling, whose spin-orbital textures can manifest deviations from the typical ones due to the spin-Rashba coupling (Sect.\ III) and can exhibit topological SC phases with orbital-driven pairing (Sect.\ IV).\@

Finally, we find that at the nodal points topological transitions for the spin-winding can occur due to the emergence of vanishing spin amplitude lines connecting the nodal points (Sect.\ IV).\@
This outcome sets the fundamental interplay between chiral and spin- or orbital winding in nodal superconductors with ISB.

\section{Topological spin-texture: single orbital model description}


\subsection{Model Hamiltonian and spin-texture}

We start by introducing a minimal model that can describe the spin-texture of the SC state due to the interplay of inversion asymmetric SO coupling and spin-triplet pairing. 
Due to ISB the pairing has mixture of spin-triplet and singlet components. 
Since the spin-singlet pairing does not affect the spin-texture, the central focus is on the consequences of the spin-triplet pair potential. 
In the superconducting state we consider the Bogoliubov-de Gennes (BdG) Hamiltonian constructed from $\hat{h}(\bm{k})$ and including both spin-singlet and triplet pairings as follows 
\begin{align}
    \hat{H}_\mathrm{BdG}(\bm{k})&=
    \begin{pmatrix}
        -\mu\hat{\sigma}_{0}+\hat{h}(\bm{k}) & \hat{\Delta}(\bm{k}) \\
        \hat{\Delta}^{\dagger}(\bm{k}) & \mu\hat{\sigma}_{0}-\hat{h}^{t}(-\bm{k})
    \end{pmatrix}, 
\end{align}%
where $\mu$ and $\hat{\Delta}(\bm{k})=i\hat{\sigma}_{y}[|\Delta_\mathrm{S}|\psi+|\Delta_\mathrm{T}|\hat{\bm{\sigma}}\cdot \bm{d}(\bm{k})]$ denote the chemical potential, and the singlet ($\psi$) and triplet order parameters ($\textbf{d}$), and the corresponding gap amplitudes $|\Delta_\mathrm{S}|$ and $|\Delta_\mathrm{T}|$, and $\hat{h}(\bm{k})$ is the normal state term 
\begin{align}
    \hat{h}(\bm{k})&=\varepsilon(\bm{k})\hat{\sigma}_{0}+\Lambda\bm{g}(\bm{k})\cdot\hat{\bm{\sigma}}, \\
    \bm{g}(\bm{k})&=(g_x(\bm{k}),g_y(\bm{k}),g_z(\bm{k})),
\end{align}%
with $\varepsilon(\bm{k})$ and $\bm{g}(\bm{k})$ being the kinetic energy and inversion asymmetry coupling, 
while $\Lambda$ denotes the strength of the ISB potential,
and $\hat{\sigma}_{i}$ $(i=0,x,y,z)$ are the Pauli matrices in spin space. Here, the $\bf{d}$-vector has the usual matrix form in terms of the components associated with the spin-triplet configurations as $\Delta_{\uparrow,\uparrow}-\Delta_{\downarrow,\downarrow}=-2d_{x}(\bm{k})$, $\Delta_{\uparrow,\uparrow}+\Delta_{\downarrow,\downarrow}=2id_{y}(\bm{k})$, and $\Delta_{\uparrow,\downarrow}+\Delta_{\downarrow,\uparrow}=2 d_{z}(\bm{k})$. 

We determine the spin polarization components by evaluating the expectation values of the related spin operators.
In the normal state, we assume that \textbf{g}-vector lies on $xy$-plane and $g_{z}(\bm{k})=0$. Then,
the eigenvalues and the corresponding eigenstates of the Hamiltonian are given by
\begin{align}
    E_{\pm}&=\varepsilon(\bm{k})\pm \Lambda\sqrt{g^{2}_{x}(\bm{k})+g^{2}_{y}(\bm{k})},\\
    |+\rangle&=
    \begin{pmatrix}
        \cos\frac{\theta}{2}\\
        e^{i\phi}\sin\frac{\theta}{2}
    \end{pmatrix}, \hspace{2mm}
    |-\rangle=
    \begin{pmatrix}
        -e^{-i\phi}\sin\frac{\theta}{2}\\
        \cos\frac{\theta}{2}
    \end{pmatrix},\notag
\end{align}%
with $\theta=\pi/2$, $\cos{\phi}=g_{x}(\bm{k})/\sqrt{g^{2}_{x}(\bm{k})+g^{2}_{y}(\bm{k})}$,
and $\sin{\phi}=g_{y}(\bm{k})/\sqrt{g^{2}_{x}(\bm{k})+g^{2}_{y}(\bm{k})}$. 
It is immediate to verify that the expectation values of the spin operators are given by
\begin{align}
    \langle\pm |\hat{S}_{x}|\pm\rangle&=\pm\frac{g_{x}(\bm{k})}{\sqrt{g^{2}_{x}(\bm{k})+g^{2}_{y}(\bm{k})}},\\
    \langle\pm |\hat{S}_{y}|\pm\rangle&=\pm\frac{g_{y}(\bm{k})}{\sqrt{g^{2}_{x}(\bm{k})+g^{2}_{y}(\bm{k})}},\\
    \langle\pm |\hat{S}_{z}|\pm\rangle&=0,
\end{align}%
where $\hat{S}_{i=x,y,z}$ are the spin operators expressed in terms of the Pauli matrices.
Thus, the $z$-component of the spin operator is zero (except that at the high symmetry points) and the in-plane $x$ and $y$-components are generally non-vanishing.

The planar structure of the spin polarization is a general consequence of the symmetry property of the model Hamiltonian. If the transformation $\hat{S}_{z} \rightarrow -\hat{S}_{z}$ is a symmetry for the quantum system upon examination, then, due to the absence of degeneracy at any $(k_x,k_y)$ different from the time reversal invariant momenta, the expectation value of the $z$-component of the spin operator is identically zero. Thus, one can focus the analysis only on the spin orientation in the $xy$-plane.

For convenience and clarity of computation, starting from the BdG Hamiltonian, one can introduce the electron component of the spin polarization within the $xy$-plane for the $m$-th excited state of the superconducting spectrum by means of the following relation
\begin{align}
    \theta^{\mathrm{SC}m}_\mathrm{S}
    &=\arg[\langle \Psi_{m}|\tilde{S}^{e}_{x}|\Psi_{m}\rangle
    +i\langle \Psi_{m}|\tilde{S}^{e}_{y}|\Psi_{m}\rangle], 
\end{align}%
where $|\Psi_{m}\rangle$ is the $m$-th eigenstate of the spectrum of the BdG Hamiltonian
and $\tilde{S}^{e}_{i=x,y,z}$ are the spin operators projected onto the electron space:
\begin{align}
    \tilde{S}^{e}_{i}&=\frac{1}{2}[1+\hat{\tau}_{3}]\otimes\hat{S}_{i},\\
    \hat{\tau}_{3}&=
    \begin{pmatrix}
        1 & 0 \\
        0 & -1
    \end{pmatrix},
\end{align}%
with the Pauli matrix $\hat{\tau}_{3}$ in Nambu space.

\begin{figure}[htbp]
    \centering
    \includegraphics[width=8.5cm]{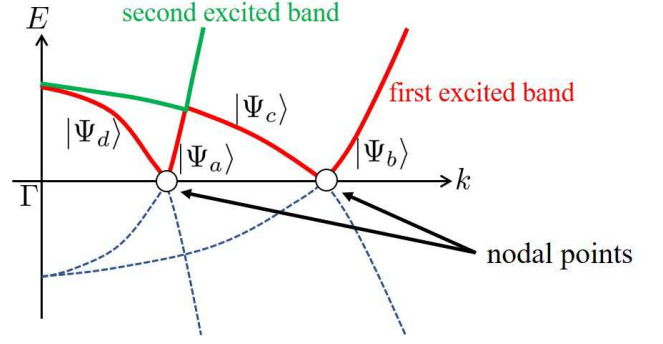}
    \caption{Schematic illustration of the relation between the eigenstates $|\Psi_{n}\rangle$ and BdG bands 
    in the case with nodal points. 
    Red (green) line is the first (second) excited band and white circle is the nodal point. 
    The same eigenstates are also plotted in Fig.\ 2 (b) with the following correspondence: 
    $|\psi_{+}\rangle=|\Psi_{a}\rangle$ and $|\psi_{-}\rangle=|\Psi_{d}\rangle$.}
    \label{Fig1}
\end{figure}

Before considering the full diagonalization of the BdG excited states, it is much instructive to consider an effective perturbation approach which allows to extract the main issues of the general behavior of the spin polarization of the superconducting excited state. 
Hence, we consider the BdG Hamiltonian by taking the first order perturbation in the pairing term,
\begin{align}
    \hat{\mathcal{H}}&=\hat{H}_{0}+\hat{H}', \\
    \hat{\mathcal{H}}|\Psi_{n}\rangle&=E_{n}|\Psi_{n}\rangle, \\
    \hat{H}_{0}|\Psi^{(0)}_{n}\rangle&=\varepsilon_{n}|\Psi^{(0)}_{n}\rangle.
\end{align}%
Here, $\hat{\mathcal{H}}$, $\hat{H}_{0}$, and $\hat{H}'$ 
correspond to the total, the unperturbed, and the perturbing Hamiltonian, respectively. 
$E_{n}$ and $|\Psi_{n}\rangle$ ($\varepsilon_{n}$ and $|\Psi^{(0)}_{n}\rangle$) 
are the eigenvalue and the corresponding eigenstate of the total Hamiltonian $\hat{\mathcal{H}}$ (the unperturbed Hamiltonian $\hat{H}_{0}$). 
Here, Fig.~\ref{Fig1} indicates the relation between the eigenstates $|\Psi_{n}\rangle$ and BdG bands. 
We assume for convenience of computation that the \textbf{g}-vector is parallel to the $z$-axis ($\bm{g}(k)=(0,0,g_{z}(k))$) 
and consider only the spin-triplet pairing ($\psi=0$). 
Then, the unperturbed and perturbed terms of the Hamiltonian at a given $k$ are written by
\begin{align}
    \hat{H}_{0}&=-\mu\hat{\sigma}_{0}\otimes\hat{\tau}_{3}+
    \begin{pmatrix}
        \hat{h}(k) & 0 \\
        0 & -\hat{h}^{t}(-k)
    \end{pmatrix}, \\
    \hat{h}(k)&=
    \begin{pmatrix}
        \varepsilon(k)+\Lambda g_{z}(k) & 0 \\
        0 & \varepsilon(k)-\Lambda g_{z}(k) 
    \end{pmatrix}, \\
    \hat{H}'&=
    \begin{pmatrix}
        0 & \hat{\Delta}(k) \\
        \hat{\Delta}^{\dagger}(k) & 0
    \end{pmatrix}, \\
    \hat{\Delta}(k)&=
    \begin{pmatrix}
        \Delta_{\uparrow,\uparrow}(k) & \Delta_{\uparrow,\downarrow}(k) \\
        \Delta_{\downarrow,\uparrow}(k) & \Delta_{\downarrow,\downarrow}(k)
    \end{pmatrix}.
\end{align}%
For the spin-triplet pairing the \textbf{d}-vector can be further expressed in terms of the polar angles $(\theta_{d},\phi_{d})$ that identify its direction in the spin space (see Fig. 2(a)) as 
\begin{align}
    \bm{d}(k)&=(d_{x}(k),d_{y}(k),d_{z}(k))\notag\\
    &=\hat{\bm{n}}(k)|\bm{d}(k)| \notag\\
    &=(\sin{\theta}_{d}\cos{\phi}_{d},\sin{\theta}_{d}\sin{\phi}_{d},\cos{\theta}_{d})|\bm{d}(k)|. \label{d_vector1}
\end{align}%
The eigenstate $|\Psi^{(0)}_{a}\rangle$ ($|\Psi^{(0)}_{c}\rangle$) corresponds to $|e,\uparrow\rangle$ ($|h,\uparrow\rangle$), 
and $|\Psi^{(0)}_{b}\rangle$ ($|\Psi^{(0)}_{d}\rangle$) is related to $|e,\downarrow\rangle$ ($|h,\downarrow\rangle$) 
where e and h are electron and hole, respectively. 
The eigenvalues and the corresponding eigenstates of the unperturbed Hamiltonian $\hat{H}_{0}$ are given by
\begin{align}
    \varepsilon_{a}(k)&=-\varepsilon_{d}(k)=\varepsilon(k)+\Lambda g_{z}(k), \\
    \varepsilon_{b}(k)&=-\varepsilon_{c}(k)=\varepsilon(k)-\Lambda g_{z}(k), \\
    |\Psi^{(0)}_{a}\rangle&=
    \begin{pmatrix}
        \hat{\alpha}_{+} \\
        \hat{0}
    \end{pmatrix},\hspace{3mm}
    |\Psi^{(0)}_{b}\rangle=
    \begin{pmatrix}
        \hat{\alpha}_{-} \\
        \hat{0}
    \end{pmatrix},\notag\\
    |\Psi^{(0)}_{c}\rangle&=
    \begin{pmatrix}
        \hat{0} \\
        \hat{\beta}_{+} 
    \end{pmatrix},\hspace{3mm}
    |\Psi^{(0)}_{d}\rangle=
    \begin{pmatrix}
        \hat{0} \\
        \hat{\beta}_{-} 
    \end{pmatrix}. \notag
\end{align}%
Here, $\hat{\alpha}_{\pm}$ and $\hat{\beta}_{\pm}$ denote the eigenstates of 
$\hat{h}(k)$ and $-\hat{h}^{t}(-k)$, 
\begin{align*}
    \hat{\alpha}_{+}&=\hat{\beta}_{+}=
    \begin{pmatrix}
        1 \\
        0
    \end{pmatrix},\hspace{3mm}
    \hat{\alpha}_{-}=\hat{\beta}_{-}=
    \begin{pmatrix}
        0 \\
        1
    \end{pmatrix}.
\end{align*}%
For the electron-like branch, 
the perturbation within the first order is zero. 
It means that the spin polarization for the electron-like branch is not modified within the first order perturbation in $|\Delta_\mathrm{T}|$, with $|\Delta_\mathrm{T}|$ being the amplitude of the spin-triplet order parameter.
On the other hand, since the eigenvalues and the corresponding eigenstates for the hole-like branch change within the first order correction, 
the spin orientation of the excited state for the hole-like branch acquires a non-trivial pattern. 
Thus, we focus on the hole-like branch of the excited state to investigate the spin-texture 
and we extract the electron component of the spin-polarization for the first excited state of the spectrum.

At this stage, by the benefit of the analytical expression of the first order eigenstates, we can calculate the spin-texture for the hole-like branch away from the Fermi level, 
that is, $|-\mu+\varepsilon(k)|\gg|\Lambda g_{z}(k)|$.
From the performed analysis, 
we can approximate the expectation values in the Appendix A as 
\begin{align}
    \langle \Psi_{c}|\tilde{S}^{e}_{x}|\Psi_{c}\rangle
    &\sim -a_s\cos{\phi}_{d}\sin{2\theta}_{d}, \notag\\
    \langle \Psi_{c}|\tilde{S}^{e}_{y}|\Psi_{c}\rangle
    &\sim -a_s\sin{\phi}_{d}\sin{2\theta}_{d}, \notag\\
    \langle \Psi_{c}|\tilde{S}^{e}_{z}|\Psi_{c}\rangle
    &\sim
    -a_s\cos{2\theta}_{d}, \label{psi3}
\end{align}%
\begin{align}
    \langle \Psi_{d}|\tilde{S}^{e}_{x}|\Psi_{d}\rangle
    &\sim a_s\cos{\phi}_{d}\sin{2\theta}_{d} \notag,\\
    \langle \Psi_{d}|\tilde{S}^{e}_{y}|\Psi_{d}\rangle
    &\sim a_s\sin{\phi}_{d}\sin{2\theta}_{d}, \notag\\
    \langle \Psi_{d}|\tilde{S}^{e}_{z}|\Psi_{d}\rangle
    &\sim
    a_s\cos{2\theta}_{d}, \label{psi4}
\end{align}%
where $a_s$ is an amplitude depending on the energy distance of the excited state 
from the Fermi level and the strength of the superconducting pairing, 
\begin{align}
    a_s=\frac{|\Delta_\mathrm{T}|^{2}|\bm{d}(k)|^{2}}{8[-\mu+\varepsilon(k)]^{2}}.
\end{align}%
\begin{figure*}
    \centering
    \includegraphics[width=15cm]{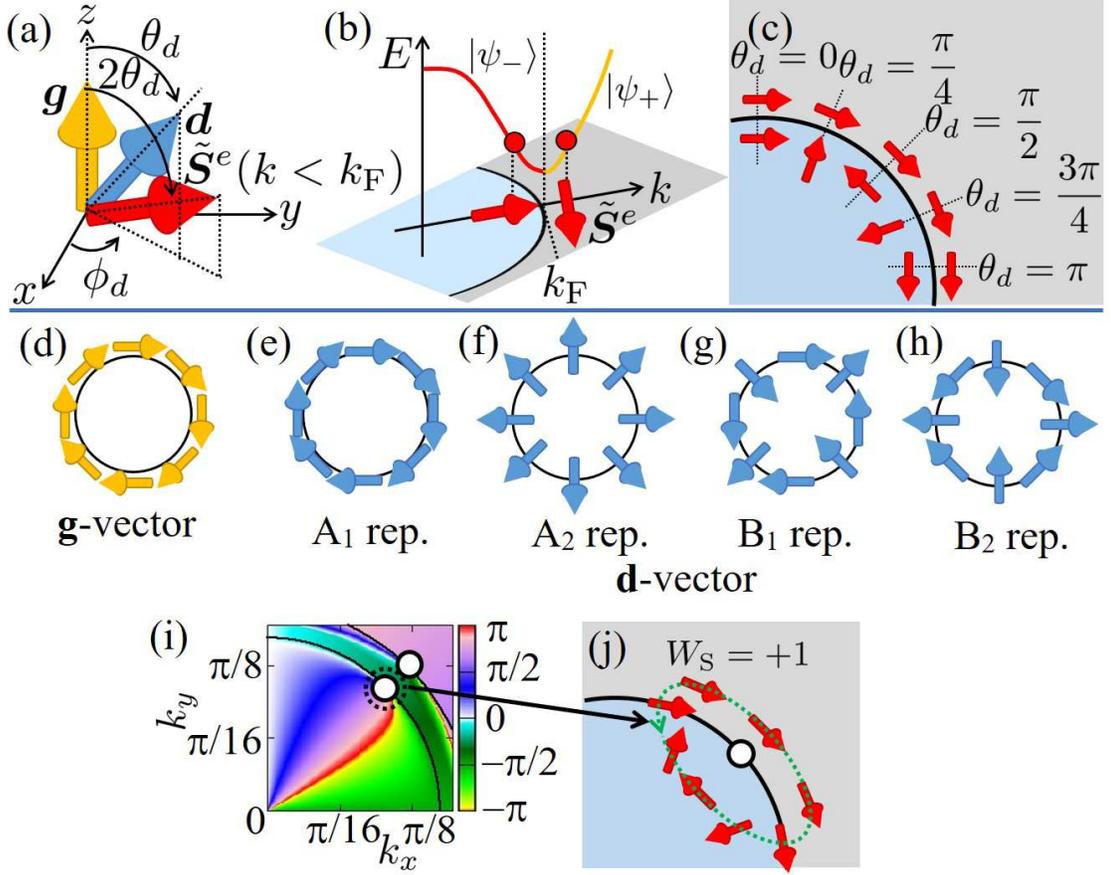}
    \caption{(a) Schematic spin-space representation of the relative orientations among the ISB \textbf{g}-vector, the spin-triplet pairing \textbf{d}-vector, and the spin direction corresponding to an excited state for $k< k_\mathrm{F}$, with $k_\mathrm{F}$ being the Fermi wave vector. $\theta_{d}$ is the polar angle between \textbf{g}- and \textbf{d}-vector and $\phi_{d}$ is the angle of the spin vector measured with respect to the in-plane $x$ direction.
    (b) Sketch of the energy dispersion along a given direction and of the spin orientation for excited states at given momentum larger (electron-like) and smaller (hole-like) than the Fermi vector in the SC state. $|\psi_{+}\rangle=|\Psi_{a}\rangle$ ($|\psi_{-}\rangle=|\Psi_{d}\rangle$) corresponds to the eigenstate for $k\ge k_\mathrm{F}$ ($k<k_\mathrm{F}$).
    Here, $\tilde{\bm{S}}^{e}$ for $k\ge k_\mathrm{F}$ is collinear to $\bm{g}$. 
    (c) Spin orientations of the excited states above and below the Fermi level at $\theta_{d}=0,\pi/4,\pi/2,3\pi/4$, and $\pi$.
    (d) Orientation of the \textbf{g}-vector and spin-vector for a Rashba-type spin-momentum coupling on the Fermi surface (black solid line).
    (e)(f)(g)(h) The \textbf{d}-vector orientation for (e) A$_1$, (f) A$_2$ (g) B$_1$, and (h) B$_2$ representations of the C$_{4v}$ point group on the Fermi surface.
    (i) Orientation of electron component of the spin-polarization for the first excited state of the B$_1$ phase in the single-band model at $\Lambda/t=8.0\times 10^{-2}$, $|\Delta_\mathrm{T}|/t=1.0\times 10^{-3}$, and $\mu/t=0.25$. The determination of the spin-polarization is done by numerical diagonalization of the superconducting model system with one band.
    Black solid line indicates the Fermi surface in the normal state and white circle is for the position of the nodal point. 
    (j) Schematic illustration of the winding spin-texture of (i) around the point node with spin-winding number $W_\mathrm{S}=+1$. }
    \label{Fig2}
\end{figure*}%
Hence, one can evaluate the character of the spin-texture from these expectation values of the spin operators. 
Then, we focus on the electron- and hole-like branch for $|\Psi_{a}\rangle$ and $|\Psi_{d}\rangle$ as shown in Fig \ref{Fig1}.

On the basis of the above observations, if $\bf{d}$- and $\bf{g}$-vectors are misaligned by an angle $\theta_{d}$ [Fig. \ref{Fig2}(a)], then the electron spin orientation corresponding to the excitations close to the Fermi level ($k_\mathrm{F}$) will manifest a distinctive pattern. 
 
This result is confirmed by full numerical determination of the BdG excited states and of the corresponding spin polarization [Fig. \ref{Fig2}(b)]. 
Indeed, one can show that for $k\ge k_\mathrm{F}$ (electron-like branch $|\Psi_{a}\rangle$) the spin orientation is collinear to the $\bf{g}$-vector while it gets rotated by an angle $2 \theta_{d}$ for $k<k_\mathrm{F}$ (hole-like branch $|\Psi_{d}\rangle$). 
Hence, a variation of the mismatch angle between $\textbf{d}$- and $\textbf{g}$-vectors along the Fermi surface can lead to a spin-texture with a general trend that is marked by an asymmetric angular dependence in the electron- and hole-branch of the low energy excitation [Fig. \ref{Fig2}(c)].  
Taking into account the configuration in Fig. \ref{Fig2}(a), one can generally demonstrate that the spin orientation for the excited state $|\psi_{+}\rangle$ at $k\ge k_\mathrm{F}$ is collinear to the ${\bf{g}}$-vector, while for $|\psi_{-}\rangle$ at $k<k_\mathrm{F}$ it depends on the angles $\theta_{d}$ and $\phi_{d}$. 
Indeed, if we define $\hat{s}^{e}_{\pm,\gamma}\equiv\langle \psi_{\pm}|\tilde{S}^{e}_{\gamma}|\psi_{\pm}\rangle$ with $\gamma=x,y,z$, the spin-vectors for electron-like branch $|\Psi_{a}\rangle=|\psi_{+}\rangle$ ($k \ge k_\mathrm{F}$) and hole-like branch $|\Psi_{d}\rangle=|\psi_{-}\rangle$ ($k<k_\mathrm{F}$) are given by
\begin{align}
    &[\hat{s}^{e}_{+,x},\hat{s}^{e}_{+,y},\hat{s}^{e}_{+,z}] \notag\\
    &\sim[\hat{\alpha}^{\dagger}_{+}\hat{S}_{x}\hat{\alpha}_{+},
    \hat{\alpha}^{\dagger}_{+}\hat{S}_{y}\hat{\alpha}_{+},
    \hat{\alpha}^{\dagger}_{+}\hat{S}_{z}\hat{\alpha}_{+}]\notag\\
    &=\left[0,0,\frac{1}{2}\right],\\
    &[\hat{s}^{e}_{-,x},\hat{s}^{e}_{-,y},\hat{s}^{e}_{-,z}] \notag\\
    &\sim [a_s \cos{\phi}_{d}\sin{2\theta}_{d}, a_s \sin{\phi}_{d} \sin{2\theta}_{d},a_s \cos{2\theta}_{d}].  
    \label{eq_spin}
\end{align}%
It is then immediate to deduce that the spin orientation is collinear to ${\textbf{g}}$ with $\theta_{d}=0$ while for perpendicularly oriented ${\textbf{g}}$- and ${\textbf{d}}$-vectors, i.e.\ $\theta_{d}=\pi/2$, the spin polarization is anti-parallel to ${\textbf{g}}$. In general, we obtain that the spin polarization lies in the same plane of ${\textbf{g}}$ and ${\textbf{d}}$ and it deviates of an angle $2\theta_{d}$ from ${\textbf{g}}$. 
Since the \textbf{d}- and \textbf{g}-vectors have the same transformation under the spin-rotation, we can generalize this result for any directions of \textbf{d}- and \textbf{g}-vectors. 
By a suitable rotation of the spin-coordinate, we can deduce the spin-texture
where \textbf{d}-vector and \textbf{g}-vector lies on $xy$-plane.


\subsection{Spin-texture at the Fermi surface}

In this subsection, we present the spin-texture evaluated at the Fermi surface, 
where $|e,\uparrow\rangle$ and $|h,\downarrow\rangle$ ($|e,\downarrow\rangle$ and $|h,\uparrow\rangle$)
are two-fold degenerate. 
We solve the BdG Hamiltonian at the Fermi surface in the case of $\varepsilon_{a}(k_\mathrm{F})=\varepsilon_{d}(k_\mathrm{F})=0$ in the basis 
$(|e\uparrow\rangle,|e\downarrow\rangle,|h\uparrow\rangle,|h\downarrow\rangle)$, 
\begin{align}
    \hat{H}(k_\mathrm{F})=
    \begin{pmatrix}
        0 & 0 & \Delta_{\uparrow,\uparrow} & \Delta_{\uparrow,\downarrow} \\
        0 & \varepsilon_{b}(k_\mathrm{F}) & \Delta_{\downarrow,\uparrow} & \Delta_{\downarrow,\downarrow} \\
        \Delta^{*}_{\uparrow,\uparrow} & \Delta^{*}_{\downarrow,\uparrow} & -\varepsilon_{b}(k_\mathrm{F}) & 0 \\
        \Delta^{*}_{\uparrow,\downarrow} & \Delta^{*}_{\downarrow,\downarrow} & 0 & 0
    \end{pmatrix},
\end{align}%
where $k_\mathrm{F}$ is the Fermi wave vector. 
We pick up the basis $(|e\uparrow\rangle,|h\downarrow\rangle)$ in this Hamiltonian  
and obtain the Hamiltonian projected onto the states $(|e\uparrow\rangle,|h\downarrow\rangle)$ near the Fermi level,
\begin{align}
    \tilde{H}(k_\mathrm{F})=
    \begin{pmatrix}
        0 & \Delta_{\uparrow,\downarrow} \\
        \Delta^{*}_{\uparrow,\downarrow} & 0
    \end{pmatrix},
\end{align}%
with $\Delta_{\uparrow,\downarrow}=|\Delta_\mathrm{T}|\cos{\theta_{d}}$.
Then, the eigenvalues are given by
\begin{align}
    E_{\pm}=\pm |\Delta_\mathrm{T}|\cos{\theta_{d}},
\end{align}%
and one of the corresponding eigenstate in the basis $(|e\uparrow\rangle,|e\downarrow\rangle,|h\uparrow\rangle,|h\downarrow\rangle)$
is for instance given by
\begin{align*}
    |+\rangle&=\frac{1}{\sqrt{2}}
    \begin{pmatrix}
        \hat{a}_{+} \\
        \hat{b}_{+}
    \end{pmatrix},\hspace{3mm}
    \hat{a}_{+}=
    \begin{pmatrix}
        1 \\
        0
    \end{pmatrix},\hspace{3mm}
    \hat{b}_{+}=
    \begin{pmatrix}
        0 \\
        1
    \end{pmatrix}.
\end{align*}%
We can obtain the eigenvalues of the electron component of the spin operator at $k_\mathrm{F}$ point, 
\begin{align}
    \langle +|\tilde{S}^{e}_{i}|+\rangle&=\frac{1}{2}\hat{a}^{\dagger}_{+}\hat{S}_{i}\hat{a}_{+},
\end{align}%
that is, 
\begin{align}
    \langle +|\tilde{S}^{e}_{x}|+\rangle&=\langle +|\tilde{S}^{e}_{y}|+\rangle=0, \\
    \langle +|\tilde{S}^{e}_{z}|+\rangle&=\frac{1}{2}\hat{a}^{\dagger}_{+}\hat{S}_{z}\hat{a}_{+}=\frac{1}{4}.
\end{align}%
Thus, the spin-texture on the Fermi surface has the same direction as that in the normal state. 

\subsection{Spin-winding in the single-orbital model with Rashba-type spin-orbit coupling}

In the single-orbital model with Rashba-type spin-orbit coupling $\textbf{g}(\bm{k})=(\sin{k_y},-\sin{k_x},0)$ [Fig.~\ref{Fig2}(d)],
the Hamiltonian in the normal state $\hat{h}(\bm{k})$ is given by
\begin{align}
    \hat{h}(\bm{k})&=
    \begin{pmatrix}
        \varepsilon(\bm{k}) & \Lambda[\sin{k_y}+i\sin{k_x}] \\
        \Lambda[\sin{k_y}-i\sin{k_x}] & \varepsilon(\bm{k})
    \end{pmatrix}.
\end{align}%
The resulting spin-polarization in the normal state rotates along the Fermi surface in the BZ and it is basically determined by the \textbf{g}-vector.
When considering the superconducting state, 
the pairing symmetry for this model is described by five irreducible representations A$_1$, A$_2$, B$_1$, B$_2$ and E of the point group C$_{4v}$
and the direction of the spin-polarization for the hole-like branch depends on these irreducible representations. 


For the A$_1$ representation, the \textbf{d}-vector is given by $\bm{d}(\bm{k})=(\sin{k_y},-\sin{k_x},0)$ [Fig. \ref{Fig2}(e)] and 
there are no nodal points in the bulk. 
Since the \textbf{d}-vector is parallel to \textbf{g}-vector in the BZ, 
that is, the relative angle between \textbf{d}- and \textbf{g}-vectors is $\theta_{d}=0,\pi$,
the direction of the spin-texture for the hole-like branch does not change from that in the normal state. 

On the other hand, the \textbf{d}-vector for the A$_2$ representation 
$\bm{d}(\bm{k})=(\sin{k_x},\sin{k_y},0)$ [Fig. \ref{Fig2}(f)] is perpendicular to the \textbf{g}-vector and
it corresponds to $\theta_{d}=\pm\pi/2$ in the BZ.\@
Then the gapless state appear and the spin-texture for the hole-like branch becomes antiparallel to that in the normal state. 

Hence, 
the spin-winding does not occur if there are no nodal points and $\theta_{d}$ does not change in the BZ.\@



In the case of B$_1$ and B$_2$ representations, 
nodal points appear along the diagonal direction and, on the $k_x$ and $k_y$-axis, respectively. 
The \textbf{d}-vectors for the B$_1$ and B$_2$ representations are given by the basis function in the point group C$_{4v}$,
$\bm{d}(\bm{k})=(\sin{k_y},\sin{k_x},0)$ [Fig. \ref{Fig2}(g)] and $\bm{d}(\bm{k})=(\sin{k_x},-\sin{k_y},0)$ [Fig. \ref{Fig2}(h)].
We explicitly determine the spin-polarization through full diagonalization of the model Hamiltonian at any momentum in the BZ for the B$_1$ representation of the C$_{4v}$ point group.
Then we look for the spin-windings in the $xy$-plane. In order to obtain the spin-vector in the $xy$-plane, that is, $\langle\psi_{\pm}|\tilde{S}^{e}_{z}|\psi_{\pm}\rangle=0$, both \textbf{d}- and \textbf{g}-vectors are in the $xy$-plane. 
Indeed, Fig.~\ref{Fig2}(i) is the orientation of the spin-polarization in the BZ for the B$_1$ representation at $\Lambda/t=8.0\times 10^{-2}$, $|\Delta_\mathrm{T}|/t=1.0\times 10^{-3}$, and $\mu/t=0.25$, and there is a two-dimensional spin-winding around the nodal point along the diagonal of the BZ [Fig. \ref{Fig2}(j)]. 

Here, we define the spin-winding number as
\begin{align}
    W_\mathrm{S}=\frac{1}{2\pi}\oint_{C} d\theta^\mathrm{SC1}_\mathrm{S}(\bm{k}),
\end{align}%
with the path of the closed loop around the nodal point $C$.
Due to the angular relation of \textbf{d}- and \textbf{g}-vectors at each $\bm{k}$ point as shown in Fig. \ref{Fig2}(c), the spin polarization can wind around the nodal point with $W_\mathrm{S}=+1$ [Fig. \ref{Fig2}(j)]. 
We note that the spin-polarization also winds around the high symmetry points in the BZ.\@ 
At this stage, it is relevant to ask whether the spin-winding always  occurs around the nodal points. By generalizing the single-band model to include higher order terms in the inversion asymmetric coupling of the type $(\sin k)^3$ or $(\sin k)^5$, we find that the spin-winding is robust and it is not affected by the modification of the \textbf{g}-vector. 
Likewise, we also obtain the spin-windings for the B$_2$ representation with nodal points on the $x$ and $y$-axis. 
The spin-winding does not always appear even if there are point nodes in the bulk.
Indeed, if the \textbf{d}-vector is parallel to the \textbf{g}-vector on the Fermi surface like a superconducting state with $d_{x^2-y^2}+f$-wave and $d_{xy}+p$-wave pairing symmetry, 
that is, $\theta_{d}=0,\pi$,
then, the spin-texture for the hole-like branch of the excited state has the same direction as that in the normal state. 
It means that the spin-texture projected onto the electron space does not wind around the point node 
and the topological spin-texture does not appear for this pairing configuration. 
Therefore, as a general remark, the presence of point nodes does not guarantee the occurrence of a spin-winding that instead requires a $\theta_{d}$ amplitude that deviates from $0$ to $\pi$. 

Below, we show that this result obtained for an effective single band model (see Appendix B) does not hold when considering a more realistic multi-orbital description of the electronic structure.


Finally, for the case of the single band model we also discuss the effects of introducing a small amplitude of the spin-singlet pairing to study the case where spin-singlet and spin-triplet pairing coexist ($|\Delta_\mathrm{S}|\neq 0$ and $|\Delta_\mathrm{T}|\neq 0$). 
One can easily verify that the spin-texture projected onto the electron space in the superconducting state also winds around the point node if the spin-singlet pairing exists.

\section{Spin-orbital texture in multi-orbital electronic systems}


\subsection{Model Hamiltonian in the normal state and definition of spin-orbital texture} 

In order to deepen the relation between spin-winding and nodal excitations beyond the single orbital description, we consider a multi-orbital model that includes both an OR term and the atomic SO coupling. 
The Hamiltonian in the basis $[(\uparrow,\downarrow)\otimes (d_{yz},d_{zx},d_{xy})]$ 
for the normal state~\cite{Khalsa2013} 
is given by 
\begin{align} 
    &\hat{H}(\bm{k})=-\mu\hat{\sigma}_{0}\otimes\hat{L}_{0}+\hat{\sigma}_0\otimes\hat{\varepsilon}(\bm{k}) \notag \\
    &+\lambda_\mathrm{SO}\sum_{i=x,y,z}\hat{\sigma}_{i}\otimes\hat{L}_{i} \notag \\
    &+\Delta_\mathrm{is}\hat{\sigma}_0\otimes\left[g_x(\bm{k})\hat{L}_{x}+g_y(\bm{k})\hat{L}_{y}\right],
\end{align}%
with $g_x(\bm{k})=-\sin{k_y}$, and $g_y(\bm{k})=\sin{k_x}$.
Here, $\hat{\varepsilon}(\bm{k})$ denotes the matrix for the kinetic energy, 
\begin{align}
    \hat{\varepsilon}(\bm{k})&=
    \begin{pmatrix}
        \varepsilon_{yz}(\bm{k}) & 0 & 0 \\
        0 & \varepsilon_{zx}(\bm{k}) &0 \\
        0 & 0 & \varepsilon_{xy}(\bm{k})
    \end{pmatrix},
\end{align}%
and the kinetic energy for each orbital is given by
\begin{align}
    \varepsilon_{yz}(\bm{k})&=2t_1(1-\cos{k_y})+2t_3(1-\cos{k_x}), \\
    \varepsilon_{zx}(\bm{k})&=2t_1(1-\cos{k_x})+2t_3(1-\cos{k_y}), \\
    \varepsilon_{xy}(\bm{k})&=4t_2-2t_2(\cos{k_x}+\cos{k_y})+\Delta_\mathrm{t},
\end{align}%
where $t_1=t=0.10$, $t_2=t$, and $t_3=0.10t$ are the hopping integral with representative amplitudes, $\Delta_\mathrm{t}=-0.50t$ is the crystal field potential associated with the breaking of the cubic symmetry. $\lambda_\mathrm{SO}$ and $\Delta_\mathrm{is}$ are 
the spin-orbit coupling constant and the inversion symmetry breaking term, respectively. 
$\hat{L}_i$ $(i=x,y,z)$ in the basis $(d_{yz},d_{zx},d_{xy})$ denotes the orbital angular momentum operator
which is projection of the $L=2$ angular momentum operator onto the $t_{2g}$ subspace, 
\begin{align*} 
    \Hat{L}_{x}=
    \begin{pmatrix}
        0 & 0 & 0 \\
        0 & 0 & i \\
        0 & -i & 0
    \end{pmatrix},
    \Hat{L}_{y}=
    \begin{pmatrix}
        0 & 0 & -i \\
        0 & 0 & 0 \\
        i & 0 & 0
    \end{pmatrix},
    \Hat{L}_{z}=
    \begin{pmatrix}
        0 & i & 0 \\
        -i & 0 & 0 \\
        0 & 0 & 0
    \end{pmatrix},
\end{align*}%
and $\hat{L}_{0}$ is a $3\times 3$ unit matrix.
In this system, there are six nondegenerate bands at $\lambda_\mathrm{SO}\neq 0$ and $\Delta_\mathrm{is}\neq 0$.

From the diagonalization of the Hamiltonian in the three-orbital model in the normal state, 
we obtain the six energy bands and the six corresponding eigenstates. 
Similar to the spin-texture, 
we define the orbital texture by the expectation values of the angular momentum operators.
Then, in order to determine the spin-orbital polarization, we calculate the six expectation values of the orbital angular momentum operator $\hat{L}_{i}$ 
and the spin operator $\hat{S}_{i}$ for the corresponding eigenstates. 
The spin-orbital polarization can be expressed in a compact notation as 
\begin{align}
    \langle \hat{A}\rangle_{n,\bm{k}}&\equiv\langle \phi_n(\bm{k}) |\hat{A}| \phi_n(\bm{k}) \rangle, \\
    \hat{A}&=\hat{L}_x,\hat{L}_y,\hat{L}_z,\hat{S}_x,\hat{S}_y,\hat{S}_z,\notag
\end{align}%
where $|\phi_n(\bm{k})\rangle$ ($n=1\sim 6$) denotes the eigenstate which corresponds to
the $n-$th energy band. 
We can define the spin-orbital texture when $\lambda_\mathrm{SO}\neq 0$ and $\Delta_\mathrm{is}\neq 0$
because the finite values of $\lambda_\mathrm{SO}$ and $\Delta_\mathrm{is}$ lift spin degeneracy.
In addition, owing to the crystal symmetry which is described by the C$_{4v}$ point group
and the TR symmetry, which is similar to the single orbital model, 
$\langle\hat{S}_z \rangle_{n,\bm{k}}$ and $\langle \hat{L}_z \rangle_{n,\bm{k}}$ 
become zero in the normal state. 
Hence, we can consider the spin- (orbital) texture in the normal state through the direction of the spin (orbital) polarization in $xy$-plane
$\theta_\mathrm{S}(\bm{k})$ ($\theta_\mathrm{L}(\bm{k})$),
\begin{align}
    \theta_\mathrm{S}(\bm{k})&=\arg[\langle \hat{S}_x \rangle_{n,\bm{k}}+i\langle\hat{S}_y\rangle_{n,\bm{k}}],\\
    \theta_\mathrm{L}(\bm{k})&=\arg[\langle \hat{L}_x \rangle_{n,\bm{k}}+i\langle\hat{L}_y\rangle_{n,\bm{k}}],
\end{align}%
for each energy band index $n$.
Moreover, we define the direction of the momentum $\bm{k}$ as $\theta_{\bm{k}}=\arg[{k_{x}+ik_{y}}]$. 


\begin{figure}
    \centering
    \includegraphics[width=8.5cm]{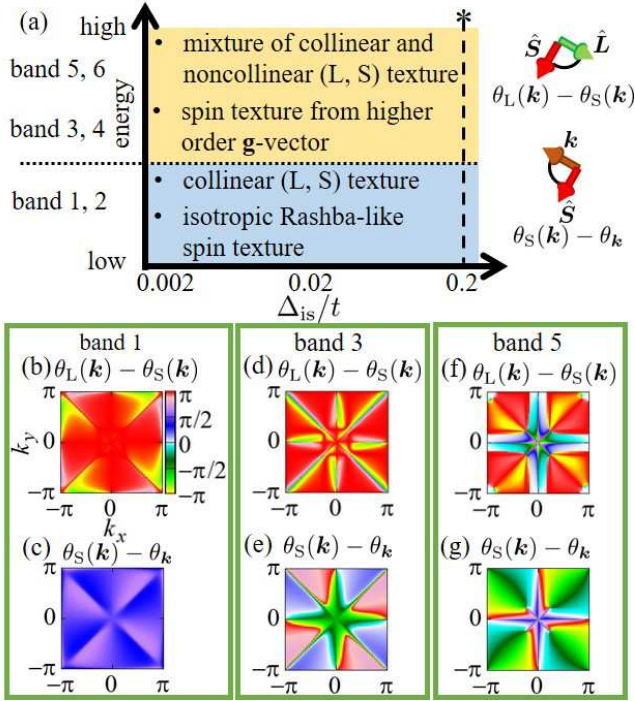}
    \caption{(a) Schematic description of the spin-orbital texture for the three-orbital model in the normal state as a function of the band index, from lowest to the highest occupied, and in terms of the OR ($\Delta_{\text{is}}$) and atomic spin-orbit ($\lambda_{\text{SO}}/t=0.10$) couplings. $\theta_\mathrm{S}(\bm{k})$, $\theta_\mathrm{L}(\bm{k})$, $\theta_{\bm{k}}$ denote the angle of the spin-, orbital- vectors and momentum $\bm{k}$ measured with respect to the $x$-axis.
    (b),(d),(f) denote the relative angle between the spin and orbital polarization for the bands 1,3,5. 
    (c),(e),(g) indicate the relative angle between the spin orientation and the momentum within the BZ.\@ The lowest occupied bands (i.e.\ 1,2) exhibit a Rashba-type spin-momentum locking. The remaining bands are marked by spin-textures with higher than the linear order in the effective $\bf{g}$-vector coupling and with a mixing of collinear and noncollinear configurations for the $\textbf{L}$ and $\textbf{S}$ angular momentum. We report only the spin-orbital pattern for the bands 1,3,5 because the others are linked to these by TR symmetry.}
    \label{Fig3}
\end{figure}%

\subsection{Spin-orbital texture in the normal state}

In this subsection, we show the spin-orbital texture in the normal state. 
A modification of the electronic amplitudes does not qualitatively alter our conclusions. 
Then, in order to evaluate the changeover of the spin-orbital texture we fix 
$\lambda_\mathrm{SO}/t=0.10$ and vary $\Delta_\mathrm{is}/t$ [Fig. \ref{Fig3}(a)] so to tune the hierarchy of the two SO interactions. 
The character of the spin- and orbital polarized states within the BZ depends on which bands are taken at the Fermi level. 
In Fig. \ref{Fig3}, we summarize the two main features of the spin-orbital textures concerning both the interrelation between the spin and orbital orientations and the spin- or orbital momentum locking. 
Firstly, due to the symmetry of the model Hamiltonian, the ISB leads to planar nonvanishing spin and orbital polarizations at any given $\bm{k}$ except for the high symmetry points with a relative angle, $\theta_\mathrm{L}(\bm{k})-\theta_\mathrm{S}(\bm{k})$, that is about uniform (collinear spin and orbital components) in the BZ for the lowest occupied bands (i.e.\ 1,2) [Fig. \ref{Fig3}(b)]. 
Here, $\theta_\mathrm{S}(\bm{k})$ ($\theta_\mathrm{L}(\bm{k})$) stands for the orientation of spin- (orbital) vector. 
The highest energy bands (i.e.\ 3,4,5,6), instead, exhibit a more intricate structure. Indeed, the spin and orbital polarizations are not anymore collinear near the high symmetry lines [Figs. \ref{Fig3}(d)(f)]. 
Such behavior is also encountered in the relative orientation of the spin polarization with respect to the direction of the momentum $\bm{k}$ set by the angle $\theta_{\bm{k}}$. 
The spin is perpendicular to the momentum (i.e.\ $\theta_\mathrm{S}(\bm{k})-\theta_{\bm{k}}\sim \pm \pi/2$) only for the lowest occupied bands (i.e.\ 1,2) [Fig. \ref{Fig3}(c)]. On the contrary, the remaining electronic states exhibit a non-isotropic spin-momentum pattern that can be accounted by the presence of higher than linear order in the direct ${\bf g}$-vector spin-momentum coupling [Figs. \ref{Fig3}(e)(g)]. This is a general behavior which is characteristic of the interplay between the OR and the atomic SO coupling (see also Sect.\ IV.\ E).

\begin{figure}
    \centering
    \includegraphics[width=8.5cm]{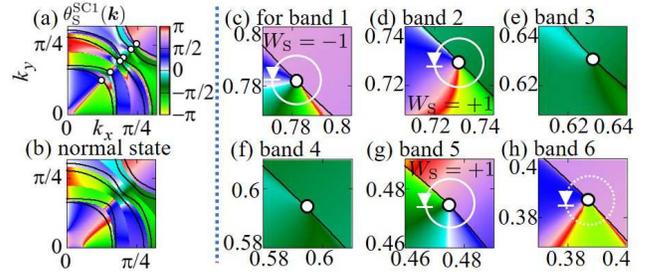}
    \caption{Spin-texture of the lowest excited states corresponding to the inter-orbital B$_1$ superconducting phase (a) and corresponding spin-pattern in the normal state including the hole branch (b) 
    at $\lambda_{\text{SO}}/t=0.10$, $\Delta_{\text{is}}/t=0.20$, $|\Delta_\mathrm{T}|/t=1.0\times 10^{-3}$, and $\mu/t=0.35$. 
    White circle indicates the nodal points.
    From (c) to (h) we zoom on the spin-texture of the superconducting excitations around the nodal points for the corresponding bands at the Fermi level from the lowest to the highest energy. The bands 1,2 and 5 exhibit spin-winding numbers around the nodal points ($W_\mathrm{S}=\pm 1$) while the excitations associated with the bands 3 and 4 have uniform spin orientation ($W_\mathrm{S}=0$), and, finally, the band 6 has an incomplete winding around the point node thus $W_\mathrm{S}=0$.}
    \label{Fig4}
\end{figure}%

\section{Topological spin-winding in nodal topological superconductors} 

\subsection{Definition of spin-orbital texture in the superconducting state}

In the superconducting state,
the BdG Hamiltonian in the three-orbital model is given by
\begin{align}
    \hat{H}_\mathrm{BdG}&=
    \begin{pmatrix}
        \hat{H}(\bm{k}) & \hat{\Delta} \\
        \hat{\Delta}^{\dagger} & -\hat{H}^{t}(-\bm{k})
    \end{pmatrix}.
\end{align}%
Here, since we focus on the local $s$-wave pairing, the superconducting order parameter associated with orbitals $\alpha$ and $\beta$ can be classified as an isotropic ($s$-wave) spin-triplet/orbital-singlet $\bm{d}^{(\alpha, \beta)}$-vector 
and $s$-wave spin-singlet/orbital-triplet with amplitude $\psi^{(\alpha, \beta)}$ or as a mixing of both configurations.
With these assumptions, one can generally describe the isotropic order parameter with spin-singlet and triplet components as 
\begin{align}
\Hat{\Delta}_{\alpha, \beta}&=
\begin{pmatrix}
    \hat{\Delta}_{\alpha\uparrow,\beta\uparrow} & \hat{\Delta}_{\alpha\uparrow,\beta\downarrow} \\
    \hat{\Delta}_{\alpha\downarrow,\beta\uparrow} & \hat{\Delta}_{\alpha\downarrow,\beta\downarrow}
\end{pmatrix},\notag\\
&=i\Hat{\sigma}_{y}\left[ \psi^{(\alpha, \beta)}+ \Hat{\bm{\sigma}}\cdot \bm{d}^{(\alpha, \beta)} \right], 
\end{align}%
with $\alpha$ and $\beta$ standing for the orbital index, and having for each channel three possible orbital flavors.
Furthermore, owing to the selected tetragonal crystal symmetry, one can achieve three different types of inter-orbital pairings. 
The spin-singlet configurations are orbital triplets and can be described by a symmetric superposition of opposite spin states in different orbitals. 
On the other hand, spin-triplet components can be expressed by means of the following $\bf{d}$-vectors: 
\begin{align}
\bm{d}^{(xy, yz)}&=\left(d^{(xy, yz)}_{x}, d^{(xy, yz)}_{y}, d^{(xy, yz)}_{z}\right), \\
\bm{d}^{(xy, zx)}&=\left(d^{(xy, zx)}_{x}, d^{(xy, zx)}_{y}, d^{(xy, zx)}_{z}\right), \\
\bm{d}^{(yz, zx)}&=\left(d^{(yz, zx)}_{x}, d^{(yz, zx)}_{y}, d^{(yz, zx)}_{z}\right).
\end{align}%
We focus on the spin-texture projected onto the electron space in the first excited state.
We define the electron component of the spin operator in the three-orbital model as 
$\tilde{S}^{e}_{i=x,y,z}$ with
\begin{align}
    \tilde{S}^{e}_{i}&=\frac{1}{2}[1+\hat{\tau}_{3}]\otimes\hat{S}_{i}\otimes\hat{L}_{0}.
\end{align}%
We then introduce the angle  $\theta^\mathrm{SC1}_\mathrm{S}(\bm{k})$ representing the direction of the spin operator in the $xy$-plane as
\begin{align}
    \theta^\mathrm{SC1}_\mathrm{S}(\bm{k})
    &=\arg[\langle \psi_{1}(\bm{k})| \tilde{S}^{e}_x|\psi_{1}(\bm{k})\rangle
    +i\langle \psi_{1}(\bm{k})|\tilde{S}^{e}_y|\psi_{1}(\bm{k})\rangle], 
\end{align}%
where $|\psi_{1}(\bm{k})\rangle$ is the eigenstate of the first excited state in the BdG Hamiltonian.



\begin{figure*}
    \centering
    \includegraphics[width=17cm]{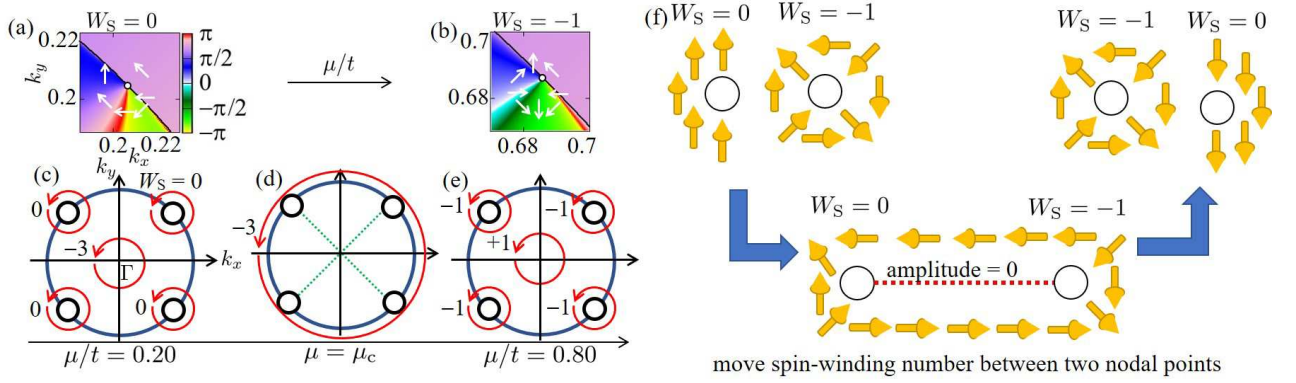}
    \caption{(a)-(b) Demonstration of topological transition of spin-winding numbers for the band 6 as a function of the chemical potential for a representative excitation branch around the nodal point at $\lambda_\mathrm{SO}/t=0.10$, $\Delta_\mathrm{is}/t=0.20$, and $|\Delta_\mathrm{T}|/t=1.0\times 10^{-3}$. 
    We set the chemical potential as (a) $\mu/t=0.35$ and (b) $\mu/t=0.80$. 
    Small circle denotes the nodal point. (c)-(e) schematically indicate the global rearrangement of the spin-winding numbers $W_\mathrm{S}$ with the occurrence at the critical amplitude of the chemical potential ($\mu_\mathrm{c}$) of lines of zero spin-amplitude (green dotted line) connecting the TR corresponding nodal points. The sum of the spin-winding numbers around the nodal points and the center of the BZ is conserved from (c) to (e).
    (f) Schematic image of the change of the spin-winding number without the deformation of superconducting gap structure. 
    Spin-winding numbers can move the line between two nodal points where the amplitude of spin polarization is zero. }
    \label{Fig5}
\end{figure*}%

\subsection{Spin-winding for the interorbital B$_1$ representation in the three-orbital model}

The starting point is to evaluate how the spin-texture changes in the SC state by focusing on the occurrence and evolution of spin-winding numbers $W_\mathrm{S}$ around the nodal points. Such feature sets the most striking changeover from the normal to the superconducting phase because the normal state does not exhibit local spin-winding close to the point nodes position. To do that we consider an inter-orbital spin-triplet/orbital-singlet/$s$-wave SC state belonging to the B$_1$ representation of the C$_{4v}$ point group~\cite{fukaya18} which is described by
\begin{align}
    \psi^{(xy,yz)}&=\psi^{(xy,zx)}=\psi^{(yz,zx)}=0, \notag\\
    \bm{d}^{(yz,zx)}&=0,\notag\\
    d^{(xy,yz)}_{z}&=d^{(xy,zx)}_{z}=d^{(xy,zx)}_{y}=d^{(xy,yz)}_{x}=0,\notag \\
    d^{(xy,zx)}_{x}&=d^{(xy,yz)}_{y}=|\Delta_\mathrm{T}|.
\end{align}%
and we set the gap amplitude for this B$_1$ representation as $|\Delta_\mathrm{T}|/t=1.0\times 10^{-3}$ 
and the chemical potential as $\mu/t=0.35$ in Fig.~\ref{Fig4}.
The spin-winding can be defined because the $z$-component of the spin-texture is zero for this B$_1$ representation.
Such configuration is well suited for our purposes because it is known~\cite{fukaya18} to be energetically favorable in a wide range of parameters 
and for this symmetry channel, the superconductor is topologically nontrivial because it exhibits nodal points along the diagonal of the BZ [Fig. \ref{Fig4}(a)]. 
Then, to single out the changeover of the SC spin-texture from that in the normal state we determine both patterns as reported in Fig. \ref{Fig4}(b).
Remarkably, its investigation for the multi-orbital topological superconductor reveals that the spin-winding is not tied to the nodal point. Indeed, for a representative set of parameters, we demonstrate that not all the excitations around the nodal position manifest a spin-winding. 

The lowest occupied bands which are well described by an effective single-band model with Rashba-type SO coupling have the same spin-winding numbers as those in the single-orbital model [Figs. \ref{Fig4}(c)-(d)]. 
On the other hand, the highest occupied bands which mainly arise from the $(d_{yz},d_{zx})$-orbitals and more significantly deviate from a Rashba-type spin-momentum locking can be employed to prove the complex topological structure of the spin-winding in the BZ [Figs. \ref{Fig4}(e)-(h)]. 
The obtained results clarify a fundamental question on the way the spin-winding around the nodal points can vary undergoing a topological transition and, in turn, affects the overall spin-pattern of the excitations. 
We point out that, if the superconductor manifests a Lifshitz-type electronic transition by merging the nodal points having opposite chiral winding numbers due to the chiral symmetry owed by the SC Hamiltonian~\cite{Yada2011,Sato2011,Brydon2011,fukaya18}, then these two nodal points have opposite spin-winding numbers and the spin-winding is also expected to disappear due to nodes annihilation and gap formation in the spectrum. 
On the other hand, it is less obvious to obtain a change of the spin-winding number without any topological modification of the nodal electronic spectrum. Hence, the investigated multi-orbital superconductor allowed to uncover a novel path for topological transitions of the spin-winding. For the band 6 corresponding to the highest occupied one, as we demonstrate in Fig. \ref{Fig5}, the spin-winding numbers for a given branch of the excitation spectra can be removed by tuning the chemical potential [see Figs. \ref{Fig5}(a)-(b)] and the transition occurs when a configuration with zero spin amplitude can be obtained in the excitation states [Figs.~\ref{Fig5}(d) and (f)]. 
Then we set the chemical potential as $\mu/t=0.35$ in Fig.~\ref{Fig5}(a) and $\mu/t=0.80$ in Fig.~\ref{Fig5}(b). 
This type of local topological transition is basically accompanied by a global change of the topological spin-winding numbers as sketched in Figs.~\ref{Fig5}(d) and (f). 
The presence of multi-orbital components in the superconductor is a fundamental requisite to achieve a quenching of the spin-momentum amplitude due to contributions of inequivalent orbital states.

\subsection{Gap amplitude dependence and comparison with the interorbital A$_1$ representation}

\begin{figure}[htbp]
    \centering
    \includegraphics[width=8.5cm]{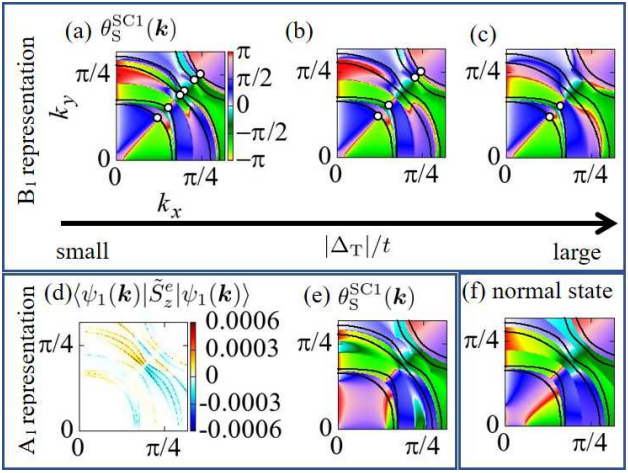}
    \caption{The direction of the electron component of the spin polarization in the superconducting state for (a)(b)(c) B$_1$ and (e) A$_1$ pairing symmetry representations in the three-orbital model at the chemical potential $\mu/t=0.35$. 
    The $z$-component of expectation value of spin in the three-orbital model 
    for the A$_1$ representation is reported in (d). 
    We set the gap amplitude 
    $|\Delta_\mathrm{T}|/t=1.0\times 10^{-3}$ for (a) and (e), $|\Delta_\mathrm{T}|/t=4.0\times 10^{-2}$ for (b), 
    and $|\Delta_\mathrm{T}|/t=0.10$ for (c). 
    (f) The direction of the spin-texture corresponding to the first excited state in the normal state 
    with $\lambda_\mathrm{SO}/t=0.10$, $\Delta_\mathrm{is}/t=0.20$, and $\mu/t=0.35$.}
    \label{Fig7}
\end{figure}

Based on the results of the previous subsection, we also consider the spin-texture as a function of the gap amplitude $|\Delta_\mathrm{T}|$ and for the interorbital A$_1$ representation. 
In Figs.~\ref{Fig7}(a), (b) and (c), we show how the electron component of the spin polarization pattern evolves
by tuning the number of point nodes through a variation of the chemical potential for the superconducting configuration belonging to the B$_1$ representation at $\mu/t=0.35$. 
We can compare these spin-textures with those of the normal state in Fig.~\ref{Fig7} (f).
With the increase of the gap amplitude $|\Delta_\mathrm{T}|$, the two nodal points with opposite spin-winding number annihilate by the Lifshitz transition as we pointed out in the previous subsection. 
Fig.~\ref{Fig7}(d) is the $z$-component of expectation value of spin operator at $\mu/t=0.35$ in the three-orbital model
for the A$_1$ representation where the interorbital pairing is described by
\begin{align}
    \psi^{(xy,yz)}&=\psi^{(xy,zx)}=\psi^{(yz,zx)}=0,\notag \\
    d^{(yz,zx)}_{x}&=d^{(yz,zx)}_{y}=0,\notag \\
    d^{(xy,yz)}_{z}&=d^{(xy,zx)}_{z}=d^{(xy,zx)}_{y}=d^{(xy,yz)}_{x}=0,\notag\\
    d^{(yz,zx)}_{z}&=-d^{(zx,yz)}_{z}=|\Delta_\mathrm{T}|,\notag\\
    d^{(xy,zx)}_{x}&=-d^{(xy,yz)}_{y}=|\Delta_\mathrm{T}|.
\end{align}%
It becomes nonzero in the BZ
owing to the spin-triplet/orbital-singlet $(d_{yz}\uparrow,d_{zx}\downarrow)$ $s$-wave pairing. 
In Fig.~\ref{Fig7}(e), we show the orientation of spin-texture in the component of $xy$-plane for the interorbital A$_1$ representation at $\mu/t=0.35$.
Then the spin-texture does not exhibit a topological structure.
Hence, one can define the topological spin winding texture for the A$_1$ representation only by specifying the axis with respect to which the spin winds.
On the other hand, we can uniquely define the in-plane spin-winding in the effective single orbital model
because both \textbf{d}-vector and \textbf{g}-vector lie on $xy$-plane as shown in the Appendix B.

\begin{figure}[htbp]
    \centering
    \includegraphics[width=8.5cm]{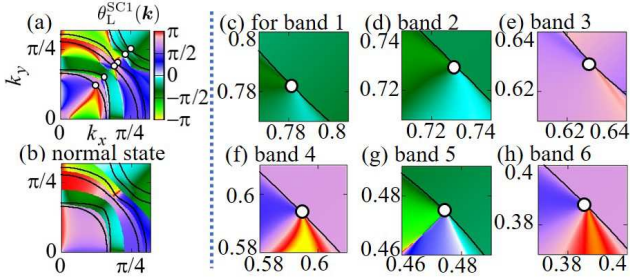}
    \caption{Orbital-texture of the lowest excited states corresponding to the inter-orbital B$_1$ superconducting phase (a) and corresponding spin-pattern in the normal state including the hole branch (b) at $\lambda_{\text{SO}}/t=0.10$, $\Delta_{\text{is}}/t=0.20$, $|\Delta_\mathrm{T}|/t=1.0\times 10^{-3}$, and $\mu/t=0.35$. 
    White circle indicates the nodal points.
    From (c) to (h) we zoom on the spin-texture of the superconducting excitations around the nodal points for the corresponding bands at the Fermi level from the lowest to the highest energy. 
    There are no orbital winding for all of bands.
    }
    \label{Fig8}
\end{figure}%

\subsection{Orbital texture for the interorbital B$_1$ representation}

After having investigated the pattern of the spin polarization, we focus on the orbital texture for the configuration of interorbital pairing with B$_1$ symmetry. 
Due to the orbital singlet nature of the superconducting state, the orbital texture does not exhibit any orbital winding around the nodal points as shown in Fig.~\ref{Fig8}. 
Although the analysis is focused on the role of spin-triplet pairing, by analogy one would get similar signatures when considering orbital-triplet with spin-singlet configurations.

\begin{figure}[htbp]
    \centering
    \includegraphics[width=6cm]{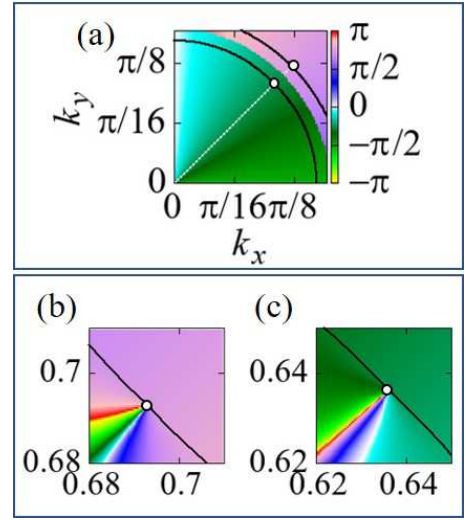}
    \caption{The direction of the spin-texture projected onto the electron space in the superconducting state 
    in the single band model (a) with \textbf{g} and \textbf{d}$_{3\mathrm{B}_{1}}$-vectors 
    and (b)(c) with \textbf{g}, \textbf{g}$_{3}$, \textbf{g}$_{5}$, 
    \textbf{d}$_{1\mathrm{B}_{1}}$, and \textbf{d}$_{3\mathrm{B}_{1}}$-vectors
    at $\Lambda_{\mathrm{R}}=8.0\times 10^{-2}$, 
    and $|\Delta_\mathrm{T}|/t=1.0\times 10^{-3}$. 
    Black solid line is Fermi surface at (a) $\mu/t=0.25$ and (b)(c) $\mu/t=0.85$.
    (b)((c)) is a magnified view around the point node on the outer (inner) Fermi surface.
    We set the parameters at 
    $\Lambda_{3\mathrm{R}}=\Lambda_{5\mathrm{R}}=0.0$, $r_{s}=0.0$, and $f=1.0$ for (a) and  
    $\Lambda_{3\mathrm{R}}=0.70$, $a_{1}=1.0$, $a_{2}=-1.0$, $\Lambda_{5\mathrm{R}}=1.0$, and $f=0.99$ for (b) and (c).}
    \label{Fig9}
\end{figure}%

\subsection{Single orbital model with higher order g-vector and d-vector with B$_1$ symmetry}

Finally, we demonstrate the consequences of higher order \textbf{g} and \textbf{d}-vectors in the single orbital model 
and compare the obtained spin polarization pattern with the spin-texture arising in the three-orbital model.
We adopt the third and fifth order \textbf{g}-vectors 
\textbf{g}$_{3}$ and \textbf{g}$_{5}$, 
and the third order \textbf{d}-vector for the B$_1$ representation \textbf{d}$_{3\mathrm{B}_{1}}$. 
Then the BdG Hamiltonian in the single orbital model is given by
\begin{align}
    \hat{H}_\mathrm{BdG}(\bm{k})&=
    \begin{pmatrix}
        \hat{h}(\bm{k}) & \hat{\Delta}(\bm{k})\\
        \hat{\Delta}^{\dagger}(\bm{k}) & -\hat{h}^{*}(-\bm{k})
    \end{pmatrix},\\
    \hat{h}(\bm{k})&=\varepsilon(\bm{k})\hat{\sigma}_{0}\notag\\
    &+[\Lambda_\mathrm{R}\bm{g}(\bm{k})+\Lambda_\mathrm{3R}\bm{g}_{3}(\bm{k})+\Lambda_\mathrm{5R}\bm{g}_{5}(\bm{k})]\cdot\hat{\bm{\sigma}}\\
    \hat{\Delta}(\bm{k})&=i\hat{\sigma}_{y}\hat{\bm{\sigma}}\cdot[(1-f)\bm{d}_{\mathrm{B}_1}(\bm{k})+f\bm{d}_{3\mathrm{B}_1}(\bm{k})],
\end{align}%
and \textbf{g}-vectors and \textbf{d}-vectors are defined as
\begin{align}
    \bm{g}(\bm{k})&=(\sin{k_y},-\sin{k_x},0),\\
    \bm{g}_{3}(\bm{k})&=\bm{g}^{(1)}_{3}(\bm{k})+\bm{g}^{(2)}_{3}(\bm{k}),\\
    \bm{g}^{(1)}_{3}(\bm{k})&=a_{1}((1-\cos{k_x})\sin{k_y},-(1-\cos{k_y})\sin{k_x},0),\\
    \bm{g}^{(2)}_{3}(\bm{k})&=a_{2}((1-\cos{k_y})\sin{k_y},-(1-\cos{k_x})\sin{k_x},0),\\
    \bm{g}_{5}(\bm{k})&=(\cos{k_x}-\cos{k_y}) \notag\\
    &\times((\cos{k_x}-1)\sin{k_y},(\cos{k_y}-1)\sin{k_x},0),\\
    \bm{d}_{\mathrm{B}_1}(\bm{k})&=|\Delta_\mathrm{T}|(\sin{k_y},\sin{k_x},0),\\
    \bm{d}_{3\mathrm{B}_1}(\bm{k})&=|\Delta_\mathrm{T}|(\cos{k_x}-\cos{k_y})\bm{g}.
\end{align}%
Here, we neglect $z$-components of \textbf{d}-vector and \textbf{g}-vector 
since we are considering the C$_{4v}$ point group. 

Fig.~\ref{Fig9}(a) reports the angular dependence of the electron component of the spin polarization 
for the first excited state of the superconducting spectrum
in the single band model assuming \textbf{g} and \textbf{d}$_{3\mathrm{B}_{1}}$-vectors at $\mu/t=0.25$. 
Here, we cannot define the spin polarization for the hole-like branch of the excited state due to $\bm{d}_{3\mathrm{B}_{1}}=(0,0,0)$
in the diagonal direction.
In addition, the spin orientation for the hole-like band has the same direction as 
that for the electron-like band
because \textbf{d}$_{3\mathrm{B}_{1}}$-vector is parallel to \textbf{g}-vector in the BZ\@.
In Figs~\ref{Fig9}(b) and (c), we show explicitly the resulting spin-texture pattern in the superconducting state
with \textbf{g}, \textbf{g}$_{3}$, \textbf{g}$_{5}$, \textbf{d}$_{1\mathrm{B}_{1}}$, and \textbf{d}$_{3\mathrm{B}_{1}}$-vectors at $\mu/t=0.85$.
In these figures, we set the ratio of \textbf{d}$_{\mathrm{B}_{1}}$ and \textbf{d}$_{3\mathrm{B}_{1}}$-vector as $f=0.99$.
The result indicates that the spin polarization winds around the point nodes only when $f$ is not equal to $1$.

\section{Conclusions}

We demonstrated that the spin-orbital texture of non-centrosymmetric superconductors with TR symmetry can unveil fundamental aspects of the pairing state. 
A mismatch of the spin (orbital) polarizations from the normal to the superconducting state can set the hallmarks of the presence of non-trivial spin- (orbital) triplet vectors. 
We clarified that the spin-winding around the nodal state can appear for the B$_1$ and B$_2$ pairings in the C$_{4v}$ point group 
because of differences of the windings of \textbf{d} and \textbf{g}-vectors. 
We note that this kind of pairings can be realized energetically when the interorbital interactions are dominant in the attractive interactions~\cite{fukaya18}. 
Furthermore, these phases can also include spin-singlet pairings ($d_{x^{2}-y^{2}}$ and $d_{xy}$-wave) owing to the ISB.\@ 

Remarkably, for pairing configurations having nodal excitations we expect to observe a local spin- (or orbital) winding which can undergo topological transitions without any change in the electronic spectrum. 
Such behavior is fundamentally tied to the degree of spin-orbital entanglement of the SC state and to a spin-momentum coupling which deviates from the Rashba-type, thus one may expect to encounter it in realistic electronic systems with ISB.\@
Our findings can have various experimentally accessible consequences. Firstly, due to the recent advancements of the application of circularly-polarized spin- and angle-resolved photoemission spectroscopy, one can employ a combination of orbital-selectivity of circularly polarized light with spin detection to directly and independently access the spin- and orbital vectors throughout the BZ. This experimental techniques are challenging and in continuous development especially when dealing with the spin detection of states with mixed orbital symmetry \cite{SpinDetect1}  
or with coherent spin phenomena in photoexcited states \cite{SpinDetect2}. 

Apart from ARPES, a weak perturbation due to an external magnetic field can lead to unconventional magnetic response with anomalous spin and orbital susceptibility.
Indeed, in nodal semimetals a changeover from large diamagnetic to paramagnetic susceptibility can be achieved when the Fermi energy moves from above to below the band crossing point.
This has been theoretically demonstrated \cite{Boiko1960} and experimentally observed \cite{Schober2012} in materials with large Rashba SO coupling. In analogy, similar effects can be also expected for the achieved nodal superconductors with the spin- and orbital textures that contribute to yield an anomalous magnetic response. 

Alternatively, one can also expect a variation of superconducting pairing symmetry due to weak external mechanical deformations~\cite{Alidoust1,Alidoust2}.
Other experimental probes might involve the measurement of the angular dependent specific heat \cite{Vorontsov2007,Sakakibara2016} or thermal conductivity \cite{Matsuda2006}. Apart from detecting the position of the nodal points under a magnetic field rotated with respect to the crystal axes, the spin-orbital structure of the nodal points can manifest into non-standard features in the specific heat or thermal conductivity behavior, e.g.\ existence of inequivalent gaps in the spin, orbital and charge excitations, spin-orbital dependent Andreev bound states within the vortex core, etc. 

\begin{acknowledgements}

This work was supported by the JSPS Core-to-Core program ``Oxide Superspin'' international network, and a JSPS KAKENHI (Grants No.\ JP15H05851, JP15H05853, JP15K21717, JP18H01176, and JP18K03538), and the project Quantox of QuantERA-NET Cofund in Quantum Technologies, implemented within the EU-H2020 Programme. 
M. Cuoco and P. Gentile acknowledge support by the project "Two-dimensional Oxides Platform for SPIN-orbitronics nanotechnology (TOPSPIN)"
funded by the MIUR Progetti di Ricerca di Rilevante Interesse Nazionale (PRIN) Bando 2017 - grant 20177SL7HC.
We acknowledge valuable comments and observations by Prof.\ E. Rashba.

\end{acknowledgements}

\section{Appendix}

\subsection{Perturbation theory in the superconducting state in the single orbital model}

In this Appendix, we show the spin-texture for the hole-like branch in the equations (\ref{psi3}) and (\ref{psi4}) analytically 
by the perturbation theory. 
The eigenstates within the first order perturbation $|\Psi^{(1)}_{n=c,d}\rangle$ 
$(|\Psi_{n}\rangle=|\Psi^{(0)}_{n}\rangle+|\Psi^{(1)}_{n}\rangle+\cdots)$ are given by
\begin{align}
    |\Psi^{(1)}_{c}\rangle&=-\frac{\hat{\alpha}^{\dagger}_{+}\hat{\Delta}\hat{\beta}_{+}}{2[-\mu+\varepsilon(k)]}
    \begin{pmatrix}
        \hat{\alpha}_{+} \\
        0
    \end{pmatrix} \notag \\
    &-\frac{\hat{\alpha}^{\dagger}_{-}\hat{\Delta}\hat{\beta}_{+}}{2[-\mu+\varepsilon(k)-\Lambda g_{z}(k)]}
    \begin{pmatrix}
        \hat{\alpha}_{-} \\
        0
    \end{pmatrix}\notag\\
    &=-\frac{\Delta_{\uparrow,\uparrow}}{2[-\mu+\varepsilon(k)]}
    \begin{pmatrix}
        \hat{\alpha}_{+} \\
        0
    \end{pmatrix}\notag \\
    &-\frac{\Delta_{\downarrow,\uparrow}}{2[-\mu+\varepsilon(k)-\Lambda g_{z}(k)]}
    \begin{pmatrix}
        \hat{\alpha}_{-} \\
        0
    \end{pmatrix},\\
    |\Psi^{(1)}_{d}\rangle&=-\frac{\hat{\alpha}^{\dagger}_{+}\hat{\Delta}\hat{\beta}_{-}}{2[-\mu+\varepsilon(k)+\Lambda g_{z}(k)]}
    \begin{pmatrix}
        \hat{\alpha}_{+} \\
        0
    \end{pmatrix} \notag\\
    &-\frac{\hat{\alpha}^{\dagger}_{-}\hat{\Delta}\hat{\beta}_{-}}{2[-\mu+\varepsilon(k)]}
    \begin{pmatrix}
        \hat{\alpha}_{-} \\
        0
    \end{pmatrix}\notag\\
    &=-\frac{\Delta_{\uparrow,\downarrow}}{2[-\mu+\varepsilon(k)+\Lambda g_{z}(k)]}
    \begin{pmatrix}
        \hat{\alpha}_{+} \\
        0
    \end{pmatrix}\notag \\
    &-\frac{\Delta_{\downarrow,\downarrow}}{2[-\mu+\varepsilon(k)]}
    \begin{pmatrix}
        \hat{\alpha}_{-} \\
        0
    \end{pmatrix}.
\end{align}%
Then, we can calculate the expectation values of the spin operators 
for the hole-branch of the first excited state of the BdG spectrum within the first order perturbation.
For $|\Psi_{c}\rangle$ and $|\Psi_{d}\rangle$ we have
\begin{align}
    &\langle \Psi_{c}|\tilde{S}^{e}_{i}|\Psi_{c}\rangle \notag \\
    &=\frac{|\Delta_{\uparrow,\uparrow}|^{2}}{4[-\mu+\varepsilon(k)]^{2}}
    \hat{\alpha}^{\dagger}_{+}\hat{S}_{i}\hat{\alpha}_{+} \notag \\
    &+\frac{|\Delta_{\downarrow,\uparrow}|^{2}}{4[-\mu+\varepsilon(k)-\Lambda g_{z}(k)]^{2}}
    \hat{\alpha}^{\dagger}_{-}\hat{S}_{i}\hat{\alpha}_{-} \notag \\
    &+\frac{\Delta^{*}_{\uparrow,\uparrow}\Delta_{\downarrow,\uparrow}}
    {4[-\mu+\varepsilon(k)][-\mu+\varepsilon(k)-\Lambda g_{z}(k)]}
    \hat{\alpha}^{\dagger}_{+}\hat{S}_{i}\hat{\alpha}_{-}\notag\\
    &+\frac{\Delta_{\uparrow,\uparrow}\Delta^{*}_{\downarrow,\uparrow}}
    {4[-\mu+\varepsilon(k)][-\mu+\varepsilon(k)-\Lambda g_{z}(k)]}
    \hat{\alpha}^{\dagger}_{-}\hat{S}_{i}\hat{\alpha}_{+}, \\
    && \nonumber
    \\
    &\langle \Psi_{d}|\tilde{S}^{e}_{i}|\Psi_{d}\rangle  \notag \\
    &=\frac{|\Delta_{\uparrow,\downarrow}|^{2}}{4[-\mu+\varepsilon(k)+\Lambda g_{z}(k)]^{2}}
    \hat{\alpha}^{\dagger}_{+}\hat{S}_{i}\hat{\alpha}_{+} \notag \\
    &+\frac{|\Delta_{\downarrow,\downarrow}|^{2}}{4[-\mu+\varepsilon(k)]^{2}}
    \hat{\alpha}^{\dagger}_{-}\hat{S}_{i}\hat{\alpha}_{-} \notag \\
    &+\frac{\Delta_{\downarrow,\downarrow}\Delta^{*}_{\uparrow,\downarrow}}
    {4[-\mu+\varepsilon(k)][-\mu+\varepsilon(k)+\Lambda g_{z}(k)]}
    \hat{\alpha}^{\dagger}_{+}\hat{S}_{i}\hat{\alpha}_{-} \notag \\
    &+\frac{\Delta^{*}_{\downarrow,\downarrow}\Delta_{\uparrow,\downarrow}}
    {4[-\mu+\varepsilon(k)][-\mu+\varepsilon(k)+\Lambda g_{z}(k)]}
    \hat{\alpha}^{\dagger}_{-}\hat{S}_{i}\hat{\alpha}_{+}.
\end{align}%
Here, $\hat{\alpha}^{\dagger}_{+}\hat{S}_{i}\hat{\alpha}_{+}$ and related terms 
denote the expectation values of the spin operators in the normal state, 
\begin{align}
    \hat{\alpha}^{\dagger}_{+}\hat{S}_{x}\hat{\alpha}_{+}&=0,\hspace{2mm}
    \hat{\alpha}^{\dagger}_{+}\hat{S}_{y}\hat{\alpha}_{+}=0,\notag\\
    \hat{\alpha}^{\dagger}_{+}\hat{S}_{z}\hat{\alpha}_{+}&=\frac{1}{2},\notag\\
    \hat{\alpha}^{\dagger}_{-}\hat{S}_{x}\hat{\alpha}_{-}&=0,\hspace{2mm}
    \hat{\alpha}^{\dagger}_{-}\hat{S}_{y}\hat{\alpha}_{-}=0,\notag\\
    \hat{\alpha}^{\dagger}_{-}\hat{S}_{z}\hat{\alpha}_{-}&=-\frac{1}{2},\notag
\end{align}%
and for the terms of the type $\hat{\alpha}^{\dagger}_{+}\hat{S}_{i}\hat{\alpha}_{-}$ we have
\begin{align}
    \hat{\alpha}^{\dagger}_{+}\hat{S}_{x}\hat{\alpha}_{-}&=\frac{1}{2},\hspace{2mm}
    \hat{\alpha}^{\dagger}_{+}\hat{S}_{y}\hat{\alpha}_{-}=-\frac{i}{2},\notag\\
    \hat{\alpha}^{\dagger}_{+}\hat{S}_{z}\hat{\alpha}_{-}&=0,\hspace{2mm}
    \hat{\alpha}^{\dagger}_{-}\hat{S}_{x}\hat{\alpha}_{+}=\frac{1}{2},\notag\\
    \hat{\alpha}^{\dagger}_{-}\hat{S}_{y}\hat{\alpha}_{+}&=\frac{i}{2},\hspace{2mm}
    \hat{\alpha}^{\dagger}_{-}\hat{S}_{z}\hat{\alpha}_{+}=0 \,.\notag
\end{align}%
Moreover, the \textbf{d}-vector given in Eq.~(\ref{d_vector1}) can be expressed in terms of the relative angle with respect to the \textbf{g}-vector providing the following quantities, 
\begin{align*}
    |\Delta_{\uparrow,\uparrow}|^{2}&=|\Delta_{\downarrow,\downarrow}|^{2}=|\Delta_\mathrm{T}|^{2}|\bm{d}(k)|^{2}\sin^{2}{\theta}_{d},\\
    |\Delta_{\uparrow,\downarrow}|^{2}&=|\Delta_{\downarrow,\uparrow}|^{2}=|\Delta_\mathrm{T}|^{2}|\bm{d}(k)|^{2}\cos^{2}{\theta}_{d},\\
    \Delta^{*}_{\uparrow,\uparrow}\Delta_{\downarrow,\uparrow}&=-|\Delta_\mathrm{T}|^{2}|\bm{d}(k)|^{2}e^{i\phi_{d}}\sin{\theta}_{d}\cos{\theta}_{d},\\
    \Delta^{*}_{\downarrow,\downarrow}\Delta_{\uparrow,\downarrow}&=|\Delta_\mathrm{T}|^{2}|\bm{d}(k)|^{2}e^{-i\phi_{d}}\sin{\theta}_{d}\cos{\theta}_{d}.
\end{align*}%
Hence, the expectation values of the spin operators projected onto the electron space for the hole-like branch of the first excited state are given by
\begin{align}
    &\langle \Psi_{c}|\tilde{S}^{e}_{x}|\Psi_{c}\rangle \notag \\
    &=-\frac{|\Delta_\mathrm{T}|^{2}|\bm{d}(k)|^{2}}{2}
    \frac{\cos{\phi}_{d}\sin{\theta}_{d}\cos{\theta}_{d}}
    {2[-\mu+\varepsilon(k)][-\mu+\varepsilon(k)-\Lambda g_{z}(k)]},\\
    &\langle \Psi_{c}|\tilde{S}^{e}_{y}|\Psi_{c}\rangle \notag \\
    &=-\frac{|\Delta_\mathrm{T}|^{2}|\bm{d}(k)|^{2}}{2}
    \frac{\sin{\phi}_{d}\sin{\theta}_{d}\cos{\theta}_{d}}
    {2[-\mu+\varepsilon(k)][-\mu+\varepsilon(k)-\Lambda g_{z}(k)]},\\
    &\langle \Psi_{c}|\tilde{S}^{e}_{z}|\Psi_{c}\rangle \notag \\
    &=\frac{|\Delta_\mathrm{T}|^{2}|\bm{d}(k)|^{2}}{2} \notag\\
    &\times\left[
        \frac{\sin^{2}{\theta}_{d}}
        {4[-\mu+\varepsilon(k)]^{2}}
        -\frac{\cos^{2}{\theta}_{d}}
        {4[-\mu+\varepsilon(k)-\Lambda g_{z}(k)]^{2}}
    \right],
\end{align}%
\begin{align}
    &\langle \Psi_{d}|\tilde{S}^{e}_{x}|\Psi_{d}\rangle \notag \\
    &=\frac{|\Delta_\mathrm{T}|^{2}|\bm{d}(k)|^{2}}{2}
    \frac{\cos{\phi}_{d}\sin{\theta}_{d}\cos{\theta}_{d}}
    {2[-\mu+\varepsilon(k)][-\mu+\varepsilon(k)+\Lambda g_{z}(k)]},\\
    &\langle \Psi_{d}|\tilde{S}^{e}_{y}|\Psi_{d}\rangle \notag \\
    &=\frac{|\Delta_\mathrm{T}|^{2}|\bm{d}(k)|^{2}}{2}
    \frac{\sin{\phi}_{d}\sin{\theta}_{d}\cos{\theta}_{d}}
    {2[-\mu+\varepsilon(k)][-\mu+\varepsilon(k)+\Lambda g_{z}(k)]},\\
    &\langle \Psi_{d}|\tilde{S}^{e}_{z}|\Psi_{d}\rangle \notag \\
    &=-\frac{|\Delta_\mathrm{T}|^{2}|\bm{d}(k)|^{2}}{2} \notag\\
    &\times\left[
        \frac{\sin^{2}{\theta}_{d}}
        {4[-\mu+\varepsilon(k)]^{2}}
        -\frac{\cos^{2}{\theta}_{d}}
        {4[-\mu+\varepsilon(k)+\Lambda g_{z}(k)]^{2}}
    \right].
\end{align}%
If $|-\mu+\varepsilon(k)|\gg|\Lambda g_{z}(k)|$, we obtain the equations (\ref{psi3}) and (\ref{psi4}).

\subsection{Derivation of single orbital effective description from the three-orbital model with atomic spin-orbit and orbital Rashba couplings}

Next, we construct the effective single orbital low-energy description near the $\Gamma$ point starting from the three-orbital model
which describes a two-dimensional non-centrosymmetric electronic system with tetragonal symmetry including the atomic spin-orbit and the orbital Rashba coupling.
The aim is to compare the spin polarization obtained in the single orbital model with that of the full multi-orbital model. 

To obtain the effective single orbital model for the normal state electronic structure near the $\Gamma$-point from the three-orbital model we employ the following perturbation scheme separating the Hamiltonian in two parts, 
\begin{align}
    \hat{H}&=\hat{H}_{0}+\hat{H}^{'},\\
    \hat{H}_{0}&=\hat{\varepsilon}(\bm{k}),\\
    \hat{H}^{'}&=\lambda_\mathrm{SO}\sum_{i=x,y,z}\hat{\sigma}_{i}\otimes\hat{L}_{i}\notag \\
    &+\Delta_\mathrm{is}\hat{\sigma}_0\otimes\left[\sin{k_x}\hat{L}_{y}-\sin{k_y}\hat{L}_{x}\right].
\end{align}%
Importantly, due to the two-dimensional confinement, the $d_{xy}$-orbital is generally well separated from the $(d_{zx},d_{yz})$-orbitals by the crystal field potential $\Delta_\mathrm{t}$.
In the normal state of $\hat{H}_{0}$, the $d_{xy}$-orbital has the lowest energy among the three orbitals and 
the energy of the $d_{xy}$-orbital is $-\Delta_{t}$ lower than $d_{yz}$ and $d_{zx}$-orbitals. 
We can consider the following process $|xy,\uparrow\rangle\rightarrow|xy,\downarrow\rangle$ 
within the second order perturbation, 
\begin{align}
    -\frac{\langle xy,\downarrow|\hat{H}^{'}|yz,\uparrow\rangle\langle yz,\uparrow|\hat{H}^{'}|xy,\uparrow\rangle}{E_{yz}-E_{xy}}
    &=\frac{i\lambda_\mathrm{SO}\Delta_\mathrm{is}\sin{k_x}}{\Delta_{t}},\\
    -\frac{\langle xy,\downarrow|\hat{H}^{'}|yz,\downarrow\rangle\langle yz,\downarrow|\hat{H}^{'}|xy,\uparrow\rangle}{E_{yz}-E_{xy}}
    &=\frac{i\lambda_\mathrm{SO}\Delta_\mathrm{is}\sin{k_x}}{\Delta_{t}},\\
    -\frac{\langle xy,\downarrow|\hat{H}^{'}|zx,\uparrow\rangle\langle zx,\uparrow|\hat{H}^{'}|xy,\uparrow\rangle}{E_{zx}-E_{xy}}
    &=-\frac{\lambda_\mathrm{SO}\Delta_\mathrm{is}\sin{k_y}}{\Delta_{t}},\\
    -\frac{\langle xy,\downarrow|\hat{H}^{'}|zx,\downarrow\rangle\langle zx,\downarrow|\hat{H}^{'}|xy,\uparrow\rangle}{E_{zx}-E_{xy}}
    &=-\frac{\lambda_\mathrm{SO}\Delta_\mathrm{is}\sin{k_y}}{\Delta_{t}},
\end{align}%
with $E_{yz}=-\Delta_{t}$ and $E_{yz}=0$.
Likewise, we can consider the following process $|xy,\downarrow\rangle\rightarrow|xy,\uparrow\rangle$ 
within the second order perturbation, 
\begin{align}
    -\frac{\langle xy,\uparrow|\hat{H}^{'}|yz,\downarrow\rangle\langle yz,\downarrow|\hat{H}^{'}|xy,\downarrow\rangle}{E_{yz}-E_{xy}}
    &=-\frac{i\lambda_\mathrm{SO}\Delta_\mathrm{is}\sin{k_x}}{\Delta_{t}},\\
    -\frac{\langle xy,\uparrow|\hat{H}^{'}|yz,\uparrow\rangle\langle yz,\uparrow|\hat{H}^{'}|xy,\downarrow\rangle}{E_{yz}-E_{xy}}
    &=-\frac{i\lambda_\mathrm{SO}\Delta_\mathrm{is}\sin{k_x}}{\Delta_{t}},\\
    -\frac{\langle xy,\uparrow|\hat{H}^{'}|zx,\downarrow\rangle\langle zx,\downarrow|\hat{H}^{'}|xy,\downarrow\rangle}{E_{zx}-E_{xy}}
    &=-\frac{\lambda_\mathrm{SO}\Delta_\mathrm{is}\sin{k_y}}{\Delta_{t}},\\
    -\frac{\langle xy,\uparrow|\hat{H}^{'}|zx,\uparrow\rangle\langle zx,\uparrow|\hat{H}^{'}|xy,\downarrow\rangle}{E_{zx}-E_{xy}}
    &=-\frac{\lambda_\mathrm{SO}\Delta_\mathrm{is}\sin{k_y}}{\Delta_{t}},
\end{align}%
then, in the subspace spanned by the states $|xy,\downarrow\rangle,|xy,\uparrow\rangle$ we obtain the effective low energy Hamiltonian in the normal state,
\begin{align} 
    \tilde{h}(\bm{k})&=
    \begin{pmatrix}
        \varepsilon_{\uparrow,\uparrow}(\bm{k}) & \varepsilon_{\uparrow,\downarrow}(\bm{k}) \\
        \varepsilon_{\downarrow,\uparrow}(\bm{k}) & \varepsilon_{\downarrow,\downarrow}(\bm{k})
    \end{pmatrix}, \\
    &=\varepsilon_{xy}(\bm{k})\hat{\sigma}_{0}
    +\Lambda_\mathrm{R}[g_{x}(\bm{k})\hat{\sigma}_{x}+g_{y}(\bm{k})\hat{\sigma}_{y}], \\
    \bm{g}(\bm{k})&=(\sin{k_y},-\sin{k_x},0),
\end{align}%
with $\Lambda_\mathrm{R}=-2\lambda_\mathrm{SO}\Delta_\mathrm{is}/\Delta_{t}$. 
Then, the elements of the Hamiltonian in the effective model $\varepsilon_{\downarrow,\uparrow}(\bm{k})$ 
and $\varepsilon_{\uparrow,\downarrow}(\bm{k})$ are derived as
\begin{align}
    \varepsilon_{\downarrow,\uparrow}(\bm{k})
    &=-\sum_{l\neq xy,\sigma}
    \frac{\langle xy,\downarrow|\hat{H}^{'}|l,\sigma\rangle\langle l,\sigma|\hat{H}^{'}|xy,\uparrow\rangle}{E_{l}-E_{xy}}\notag \\
    &=\Lambda_\mathrm{R}[\sin{k_y}-i\sin{k_x}],\\
    \varepsilon_{\uparrow,\downarrow}(\bm{k})
    &=-\sum_{l\neq xy,\sigma}
    \frac{\langle xy,\uparrow|\hat{H}^{'}|l,\sigma\rangle\langle l,\sigma|\hat{H}^{'}|xy,\downarrow\rangle}{E_{l}-E_{xy}}\notag \\
    &=\Lambda_\mathrm{R}[\sin{k_y}+i\sin{k_x}],
\end{align}%
where $l=yz,zx,xy$ are the indices of the $d$-orbitals and $\sigma=\uparrow,\downarrow$ indicate the spin polarizations.
On the other hand, because there are no processes $|xy,\uparrow\rangle\rightarrow|xy,\uparrow\rangle$ 
and $|xy,\downarrow\rangle\rightarrow|xy,\downarrow\rangle$ within the second order perturbation, 
we obtain
\begin{align}
    \varepsilon_{\uparrow,\uparrow}(\bm{k})=\varepsilon_{\downarrow,\downarrow}(\bm{k})=\varepsilon_{xy}(\bm{k}).
\end{align}%
The effective low energy description is then expressed as a single-orbital model with a Rashba-type spin-orbit coupling. 

In a similar fashion, for the superconducting state one consider the following perturbation scheme,
\begin{align}
    \hat{H}_\mathrm{BdG}&=\hat{H}^{0}_\mathrm{BdG}+\hat{H}^{'}_\mathrm{BdG},\\
    \hat{H}^{0}_\mathrm{BdG}&=
    \begin{pmatrix}
        \hat{\varepsilon}(\bm{k}) & 0 \\
        0 & -\hat{\varepsilon}(-\bm{k})
    \end{pmatrix},\\
    \hat{H}^{'}_\mathrm{BdG}&=
    \begin{pmatrix}
        \hat{H}_\mathrm{SO} & 0 \\
        0 & -\hat{H}^{t}_\mathrm{SO}
    \end{pmatrix}
    +
    \begin{pmatrix}
        \hat{H}_\mathrm{is}(\bm{k}) & 0 \\
        0 & -\hat{H}^{t}_\mathrm{is}(-\bm{k})
    \end{pmatrix}\notag  \\
    &+
    \begin{pmatrix}
        0 & \hat{\Delta} \\
        \hat{\Delta}^{\dagger} & 0
    \end{pmatrix},\\
    \hat{H}_\mathrm{SO}&=\lambda_\mathrm{SO}\sum_{i=x,y,z}\hat{\sigma}_{i}\otimes\hat{L}_{i},\\
    \hat{H}_\mathrm{is}(\bm{k})&=\Delta_\mathrm{is}\hat{\sigma}_0\otimes\left[\sin{k_x}\hat{L}_{y}-\sin{k_y}\hat{L}_{x}\right].
\end{align}%
The energy of $|yz,\sigma,h\rangle$ and $|zx,\sigma,h\rangle$ states is $-\Delta_{t}$ lower 
than that of $|xy,\sigma,h\rangle$ state. 
Here $\sigma=\uparrow,\downarrow$ denotes the spin of the electron and hole. 
The effective BdG Hamiltonian from the three-orbital model is given by
\begin{align} 
    \tilde{H}_\mathrm{BdG}&=
    \begin{pmatrix}
        \tilde{h}(\bm{k}) & \tilde{\Delta} \\
        \tilde{\Delta}^{\dagger} & -\tilde{h}^{t}(-\bm{k})
    \end{pmatrix}, \\
    \tilde{\Delta}(\bm{k})&=
    \begin{pmatrix}
        \Delta_{\uparrow,\uparrow} & \Delta_{\uparrow,\downarrow} \\
        \Delta_{\downarrow,\uparrow} & \Delta_{\downarrow,\downarrow}
    \end{pmatrix}, \notag\\
    &=
    \begin{pmatrix}
        \Delta_{\uparrow,\uparrow} & \Delta^\mathrm{S}_{\uparrow,\downarrow}+\Delta^\mathrm{T}_{\uparrow,\downarrow} \\
        \Delta^\mathrm{S}_{\downarrow,\uparrow}+\Delta^\mathrm{T}_{\downarrow,\uparrow} & \Delta_{\downarrow,\downarrow}
    \end{pmatrix}.
\end{align}%
This effective Hamiltonian can be obtained by the following processes within the second order perturbation,
\begin{align}
    &-\frac{\langle xy,\uparrow,e|\hat{H}^{'}_\mathrm{BdG}|yz,\uparrow,h\rangle
    \langle yz,\uparrow,h|\hat{H}^{'}_\mathrm{BdG}|xy,\uparrow,h\rangle}{E_{yz,h}-E_{xy}}\notag \\
    &=\frac{i\Delta_\mathrm{is}\Delta_{xy\uparrow,yz\uparrow}\sin{k_x}}{\Delta_{t}},\\
    &-\frac{\langle xy,\uparrow,e|\hat{H}^{'}_\mathrm{BdG}|yz,\uparrow,e\rangle
    \langle yz,\uparrow,e|\hat{H}^{'}_\mathrm{BdG}|xy,\uparrow,h\rangle}{E_{yz,e}-E_{xy}}\notag \\
    &=\frac{i\Delta_\mathrm{is}\Delta_{xy\uparrow,yz\uparrow}\sin{k_x}}{\Delta_{t}},\\
    &-\frac{\langle xy,\uparrow,e|\hat{H}^{'}_\mathrm{BdG}|zx,\uparrow,h\rangle
    \langle zx,\uparrow,h|\hat{H}^{'}_\mathrm{BdG}|xy,\uparrow,h\rangle}{E_{yz,h}-E_{xy}}\notag \\
    &=\frac{i\Delta_\mathrm{is}\Delta_{xy\uparrow,zx\uparrow}\sin{k_y}}{\Delta_{t}},\\
    &-\frac{\langle xy,\uparrow,e|\hat{H}^{'}_\mathrm{BdG}|zx,\uparrow,e\rangle
    \langle zx,\uparrow,e|\hat{H}^{'}_\mathrm{BdG}|xy,\uparrow,h\rangle}{E_{yz,e}-E_{xy}}\notag \\
    &=\frac{i\Delta_\mathrm{is}\Delta_{xy\uparrow,zx\uparrow}\sin{k_y}}{\Delta_{t}},
\end{align}%
\begin{align}
    &-\frac{\langle xy,\downarrow,e|\hat{H}^{'}_\mathrm{BdG}|yz,\downarrow,h\rangle
    \langle yz,\downarrow,h|\hat{H}^{'}_\mathrm{BdG}|xy,\uparrow,h\rangle}{E_{yz,h}-E_{xy}}\notag \\
    &=\frac{-\lambda_\mathrm{SO}\Delta_{xy\downarrow,yz\downarrow}}{\Delta_{t}},\\
    &-\frac{\langle xy,\downarrow,e|\hat{H}^{'}_\mathrm{BdG}|yz,\downarrow,e\rangle
    \langle yz,\downarrow,e|\hat{H}^{'}_\mathrm{BdG}|xy,\uparrow,h\rangle}{E_{yz,e}-E_{xy}}\notag \\
    &=\frac{\lambda_\mathrm{SO}\Delta_{xy\uparrow,yz\uparrow}}{-\Delta_{t}},\\
    &-\frac{\langle xy,\downarrow,e|\hat{H}^{'}_\mathrm{BdG}|zx,\downarrow,h\rangle
    \langle zx,\downarrow,h|\hat{H}^{'}_\mathrm{BdG}|xy,\uparrow,h\rangle}{E_{zx,h}-E_{xy}}\notag \\
    &=\frac{i\lambda_\mathrm{SO}\Delta_{xy\downarrow,zx\downarrow}}{\Delta_{t}},\\
    &-\frac{\langle xy,\downarrow,e|\hat{H}^{'}_\mathrm{BdG}|zx,\downarrow,e\rangle
    \langle zx,\downarrow,e|\hat{H}^{'}_\mathrm{BdG}|xy,\uparrow,h\rangle}{E_{zx,e}-E_{xy}}\notag \\
    &=-\frac{i\lambda_\mathrm{SO}\Delta_{xy\uparrow,zx\uparrow}}{\Delta_{t}}, 
\end{align}%
\begin{align}
    &-\frac{\langle xy,\downarrow,e|\hat{H}^{'}_\mathrm{BdG}|yz,\downarrow,e\rangle
    \langle yz,\downarrow,e|\hat{H}^{'}_\mathrm{BdG}|xy,\uparrow,h\rangle}{E_{yz,e}-E_{xy}}\notag \\
    &=\frac{-i\Delta_\mathrm{is}\Delta_{xy\uparrow,yz\downarrow}\sin{k_x}}{\Delta_{t}},\\
    &-\frac{\langle xy,\downarrow,e|\hat{H}^{'}_\mathrm{BdG}|yz,\uparrow,h\rangle
    \langle yz,\uparrow,h|\hat{H}^{'}_\mathrm{BdG}|xy,\uparrow,h\rangle}{E_{yz,h}-E_{xy}}\notag \\
    &=\frac{i\Delta_\mathrm{is}\Delta_{xy\uparrow,yz\downarrow}\sin{k_x}}{-\Delta_{t}},\\
    &-\frac{\langle xy,\downarrow,e|\hat{H}^{'}_\mathrm{BdG}|zx,\downarrow,e\rangle
    \langle zx,\downarrow,e|\hat{H}^{'}_\mathrm{BdG}|xy,\uparrow,h\rangle}{E_{zx,e}-E_{xy}}\notag \\
    &=\frac{-i\Delta_\mathrm{is}\Delta_{xy\uparrow,zx\downarrow}\sin{k_y}}{\Delta_{t}},\\
    &-\frac{\langle xy,\downarrow,e|\hat{H}^{'}_\mathrm{BdG}|zx,\uparrow,h\rangle
    \langle zx,\uparrow,h|\hat{H}^{'}_\mathrm{BdG}|xy,\uparrow,h\rangle}{E_{zx,h}-E_{xy}}\notag \\
    &=-\frac{i\Delta_\mathrm{is}\Delta_{xy\uparrow,zx\downarrow}\sin{k_y}}{\Delta_{t}}, 
\end{align}%
\begin{align}
    &-\frac{\langle xy,\uparrow,e|\hat{H}^{'}_\mathrm{BdG}|yz,\uparrow,h\rangle
    \langle yz,\uparrow,h|\hat{H}^{'}_\mathrm{BdG}|xy,\downarrow,h\rangle}{E_{yz,h}-E_{xy}}\notag \\
    &=\frac{\lambda_\mathrm{SO}\Delta_{xy\uparrow,yz\uparrow}}{\Delta_{t}},\\
    &-\frac{\langle xy,\uparrow,e|\hat{H}^{'}_\mathrm{BdG}|yz,\downarrow,e\rangle
    \langle yz,\downarrow,e|\hat{H}^{'}_\mathrm{BdG}|xy,\downarrow,h\rangle}{E_{yz,e}-E_{xy}}\notag \\
    &=\frac{-\lambda_\mathrm{SO}\Delta_{xy\downarrow,yz\downarrow}}{-\Delta_{t}},\\
    &-\frac{\langle xy,\uparrow,e|\hat{H}^{'}_\mathrm{BdG}|zx,\uparrow,h\rangle
    \langle zx,\uparrow,h|\hat{H}^{'}_\mathrm{BdG}|xy,\downarrow,h\rangle}{E_{zx,h}-E_{xy}}\notag \\
    &=\frac{i\lambda_\mathrm{SO}\Delta_{xy\uparrow,zx\uparrow}}{\Delta_{t}},\\
    &-\frac{\langle xy,\uparrow,e|\hat{H}^{'}_\mathrm{BdG}|zx,\downarrow,e\rangle
    \langle zx,\downarrow,e|\hat{H}^{'}_\mathrm{BdG}|xy,\downarrow,h\rangle}{E_{zx,e}-E_{xy}}\notag \\
    &=\frac{i\lambda_\mathrm{SO}\Delta_{xy\downarrow,yz\downarrow}}{-\Delta_{t}}, 
\end{align}%
\begin{align}
    &-\frac{\langle xy,\uparrow,e|\hat{H}^{'}_\mathrm{BdG}|yz,\uparrow,e\rangle
    \langle yz,\uparrow,e|\hat{H}^{'}_\mathrm{BdG}|xy,\downarrow,h\rangle}{E_{yz,e}-E_{xy}}\notag \\
    &=-\frac{i\Delta_\mathrm{is}\Delta_{xy\downarrow,yz\uparrow}\sin{k_x}}{\Delta_{t}},\\
    &-\frac{\langle xy,\uparrow,e|\hat{H}^{'}_\mathrm{BdG}|yz,\downarrow,h\rangle
    \langle yz,\downarrow,h|\hat{H}^{'}_\mathrm{BdG}|xy,\downarrow,h\rangle}{E_{yz,h}-E_{xy}}\notag \\
    &=-\frac{i\Delta_\mathrm{is}\Delta_{xy\downarrow,yz\uparrow}\sin{k_x}}{\Delta_{t}},\\
    &-\frac{\langle xy,\uparrow,e|\hat{H}^{'}_\mathrm{BdG}|zx,\uparrow,e\rangle
    \langle zx,\uparrow,e|\hat{H}^{'}_\mathrm{BdG}|xy,\downarrow,h\rangle}{E_{zx,e}-E_{xy}}\notag \\
    &=-\frac{i\Delta_\mathrm{is}\Delta_{xy\downarrow,zx\uparrow}\sin{k_x}}{\Delta_{t}},\\
    &-\frac{\langle xy,\uparrow,e|\hat{H}^{'}_\mathrm{BdG}|zx,\downarrow,h\rangle
    \langle zx,\downarrow,h|\hat{H}^{'}_\mathrm{BdG}|xy,\downarrow,h\rangle}{E_{zx,h}-E_{xy}}\notag \\
    &=-\frac{i\Delta_\mathrm{is}\Delta_{xy\downarrow,zx\uparrow}\sin{k_x}}{\Delta_{t}}, 
\end{align}%
\begin{align}
    &-\frac{\langle xy,\downarrow,e|\hat{H}^{'}_\mathrm{BdG}|yz,\downarrow,h\rangle
    \langle yz,\downarrow,h|\hat{H}^{'}_\mathrm{BdG}|xy,\downarrow,h\rangle}{E_{yz,h}-E_{xy}}\notag \\
    &=\frac{i\Delta_\mathrm{is}\Delta_{xy\downarrow,yz\downarrow}\sin{k_x}}{\Delta_{t}},\\
    &-\frac{\langle xy,\downarrow,e|\hat{H}^{'}_\mathrm{BdG}|yz,\downarrow,e\rangle
    \langle yz,\downarrow,e|\hat{H}^{'}_\mathrm{BdG}|xy,\downarrow,h\rangle}{E_{yz,e}-E_{xy}}\notag \\
    &=\frac{i\Delta_\mathrm{is}\Delta_{xy\downarrow,yz\downarrow}\sin{k_x}}{\Delta_{t}},\\
    &-\frac{\langle xy,\downarrow,e|\hat{H}^{'}_\mathrm{BdG}|zx,\downarrow,h\rangle
    \langle zx,\downarrow,h|\hat{H}^{'}_\mathrm{BdG}|xy,\downarrow,h\rangle}{E_{zx,h}-E_{xy}}\notag \\
    &=\frac{i\Delta_\mathrm{is}\Delta_{xy\downarrow,zx\downarrow}\sin{k_x}}{\Delta_{t}},\\
    &-\frac{\langle xy,\downarrow,e|\hat{H}^{'}_\mathrm{BdG}|zx,\downarrow,e\rangle
    \langle zx,\downarrow,e|\hat{H}^{'}_\mathrm{BdG}|xy,\downarrow,h\rangle}{E_{zx,e}-E_{xy}}\notag \\
    &=\frac{i\Delta_\mathrm{is}\Delta_{xy\downarrow,zx\downarrow}\sin{k_x}}{\Delta_{t}},
\end{align}%
with $E_{yz,h}=E_{zx,h}=\Delta_{t}$ and $E_{yz,e}=E_{zx,e}=-\Delta_{t}$.
Then, we obtain the elements of the BdG Hamiltonian in the effective single-orbital description for the $d_{xy}$-band 
$\Delta_{\uparrow,\uparrow}$, $\Delta_{\uparrow,\downarrow}$,
$\Delta_{\downarrow,\uparrow}$, and $\Delta_{\downarrow,\downarrow}$,
\begin{align}
    &\Delta_{\uparrow,\uparrow}\notag \\
    &=-\sum_{l\neq xy,\sigma,\tau}
    \frac{\langle xy,\uparrow,e|\hat{H}^{'}_\mathrm{BdG}|l,\sigma,\tau\rangle\langle l,\sigma,\tau|\hat{H}^{'}_\mathrm{BdG}|xy,\uparrow,h\rangle}
    {E_{l,\tau}-E_{xy}}\notag \\
    &=\frac{2i\Delta_\mathrm{is}}{\Delta_{t}}
    \left[\Delta_{xy\uparrow,yz\uparrow}\sin{k_x}+\Delta_{xy\uparrow,zx\uparrow}\sin{k_y}\right],
\end{align}%
\begin{align}
    &\Delta_{\downarrow,\uparrow}\notag \\
    &=-\sum_{l\neq xy,\sigma,\tau}
    \frac{\langle xy,\downarrow,e|\hat{H}^{'}_\mathrm{BdG}|l,\sigma,\tau\rangle\langle l,\sigma,\tau|\hat{H}^{'}_\mathrm{BdG}|xy,\uparrow,h\rangle}
    {E_{l,\tau}-E_{xy}}\notag \\
    &=\Delta^\mathrm{S}_{\downarrow,\uparrow}+\Delta^\mathrm{T}_{\downarrow,\uparrow}.
\end{align}%
\begin{align}
    \Delta^\mathrm{S}_{\downarrow,\uparrow}
    &=-\frac{i\lambda_\mathrm{SO}}{\Delta_{t}}
    [\Delta_{xy\uparrow,yz\uparrow}+\Delta_{xy\downarrow,yz\downarrow}\notag  \\
    &+i\Delta_{xy\uparrow,zx\uparrow}-i\Delta_{xy\downarrow,zx\downarrow}],\\
    \Delta^\mathrm{T}_{\downarrow,\uparrow}
    &=-\frac{2i\Delta_\mathrm{is}}{\Delta_{t}}
    \left[\Delta_{xy\uparrow,yz\downarrow}\sin{k_x}+\Delta_{xy\uparrow,zx\downarrow}\sin{k_y}\right].
\end{align}%
\begin{align}
    &\Delta_{\uparrow,\downarrow}\notag \\
    &=-\sum_{l\neq xy,\sigma,\tau}
    \frac{\langle xy,\uparrow,e|\hat{H}^{'}_\mathrm{BdG}|l,\sigma,\tau\rangle\langle l,\sigma,\tau|\hat{H}^{'}_\mathrm{BdG}|xy,\downarrow,h\rangle}
    {E_{l,\tau}-E_{xy}}\notag \\
    &=\Delta^\mathrm{S}_{\uparrow,\downarrow}+\Delta^\mathrm{T}_{\uparrow,\downarrow},
\end{align}%
\begin{align}
    \Delta^\mathrm{S}_{\uparrow,\downarrow}
    &=\frac{i\lambda_\mathrm{SO}}{\Delta_{t}}
    [\Delta_{xy\uparrow,yz\uparrow}+\Delta_{xy\downarrow,yz\downarrow}\notag \\
    &+i\Delta_{xy\uparrow,zx\uparrow}-i\Delta_{xy\downarrow,zx\downarrow}],\\
    \Delta^\mathrm{T}_{\uparrow,\downarrow}
    &=-\frac{2i\Delta_\mathrm{is}}{\Delta_{t}}
    \left[\Delta_{xy\downarrow,yz\uparrow}\sin{k_x}+\Delta_{xy\downarrow,zx\uparrow}\sin{k_y}\right],
\end{align}%
\begin{align}
    &\Delta_{\downarrow,\downarrow}\notag \\
    &=-\sum_{l\neq xy,\sigma,\tau}
    \frac{\langle xy,\downarrow,e|\hat{H}^{'}_\mathrm{BdG}|l,\sigma,\tau\rangle\langle l,\sigma,\tau|\hat{H}^{'}_\mathrm{BdG}|xy,\downarrow,h\rangle}
    {E_{l,\tau}-E_{xy}}\notag \\
    &=\frac{2i\Delta_\mathrm{is}}{\Delta_{t}}
    \left[\Delta_{xy\downarrow,yz\downarrow}\sin{k_x}+\Delta_{xy\downarrow,zx\downarrow}\sin{k_y}\right],
\end{align}%
where $\tau=e,h$ is the index of electron and hole space and superscript S and T are the spin-singlet and spin-triplet pairing 
in the $(\uparrow,\downarrow)$ sector of the gap function, respectively.

For the B$_1$ representation in the point group C$_{4v}$, the gap function in the effective model is
\begin{align}
    \Delta_{\uparrow,\uparrow}&=|\Delta^\mathrm{T}_0|(-\sin{k_y}+i\sin{k_x}),\\
    \Delta_{\uparrow,\downarrow}&=\Delta_{\downarrow,\uparrow}=0, \\
    \Delta_{\downarrow,\downarrow}&=|\Delta^\mathrm{T}_0|(\sin{k_y}+i\sin{k_x}),
\end{align}%
with $|\Delta^\mathrm{T}_0|=|2i |\Delta_\mathrm{T}|d^{(xy,yz)}_{y}|/\Delta_{t}$.
Hence, we can obtain the \textbf{d}-vector for the B$_1$ representation in the effective model: 
\begin{align}
    d_{x}(\bm{k})&=\frac{1}{2}[\Delta_{\downarrow,\downarrow}-\Delta_{\uparrow,\uparrow}]=|\Delta^\mathrm{T}_0|\sin{k_y},\notag  \\
    d_{y}(\bm{k})&=\frac{1}{2i}[\Delta_{\uparrow,\uparrow}+\Delta_{\downarrow,\downarrow}]=|\Delta^\mathrm{T}_0|\sin{k_x}, \notag \\
    d_{z}(\bm{k})&=\Delta^\mathrm{T}_{\uparrow,\downarrow}=0.
\end{align}%
It corresponds to the base functions of the spin-triplet pairing for the B$_1$ representation in the C$_{4v}$ point group. 
On the other hand, we obtain the gap function in the effective model for the A$_1$ representation, 
\begin{align}
    \Delta_{\uparrow,\uparrow}&=|\Delta^\mathrm{T}_0|(\sin{k_y}+i\sin{k_x}),\\
    \Delta^\mathrm{T}_{\uparrow,\downarrow}&=\Delta^\mathrm{T}_{\downarrow,\uparrow}=0, \\
    \Delta^\mathrm{S}_{\uparrow,\downarrow}&=-\Delta^\mathrm{S}_{\downarrow,\uparrow}=|\Delta^\mathrm{S}_0|, \\
    \Delta_{\downarrow,\downarrow}&=|\Delta^\mathrm{T}_0|(-\sin{k_y}+i\sin{k_x}),\\
    |\Delta^\mathrm{S}_0|&=-\frac{4\lambda_\mathrm{SO}d^{(xy,yz)}_{y}}{\Delta_{t}}.
\end{align}%
Therefore, the pairings for the A$_1$ representation in the effective single-orbital model are
\begin{align}
    \psi&=\Delta^\mathrm{S}_{\uparrow,\downarrow}=|\Delta^\mathrm{S}_0|,\notag \\
    d_{x}(\bm{k})&=\frac{1}{2}[\Delta_{\downarrow,\downarrow}-\Delta_{\uparrow,\uparrow}]=-|\Delta^\mathrm{T}_0|\sin{k_y}, \notag \\
    d_{y}(\bm{k})&=\frac{1}{2i}[\Delta_{\uparrow,\uparrow}+\Delta_{\downarrow,\downarrow}]=|\Delta^\mathrm{T}_0|\sin{k_x}, \notag \\
    d_{z}(\bm{k})&=\Delta^\mathrm{T}_{\uparrow,\downarrow}=0.
\end{align}%
This \textbf{d}-vector corresponds to the $s+p$-wave for the A$_1$ representation in the C$_{4v}$ point group
and it is parallel to the \textbf{g}-vector in the BZ.\@ 



\end{document}